\documentclass[a4paper,12pt]{article}

\usepackage{theorem}
\usepackage{amsfonts}

\theoremstyle{change}
\newtheorem{thm}{Theorem.}[section]

\newtheorem{cor}[thm]{Corollary.}
\newtheorem{prop}[thm]{Proposition.}
\theorembodyfont{\rmfamily}
\newtheorem{dfn}[thm]{Definition.}
\newtheorem{expl}[thm]{Example.}
\newtheorem{expls}[thm]{Examples.}

\newtheorem{rem}[thm]{Remark.}

\def\Box{\hbox{\raisebox{0.0em}{\rlap{$\sqcap$}}\kern0em%
            \raisebox{-0.0em}{$\sqcup$}} } 
\newenvironment{proof}{{\it Proof. }}{\hfill$\Box$\vspace{0.5cm}}
\def\bepf{\begin{proof}}
\def\epf{\end{proof}}

\def\fct#1{\mathop{\rm #1}}	
\def\CHI{\mbox{\large $\chi$}}

\def\cov{\fct{cov}}

\def\dim{\fct{dim}}

\def\im{\fct{Im}}

\def\re{\fct{Re}}

\def\spec{\fct{Spec}}

\def\tr{\fct{tr}}

\def\newl{\hfill\break}				
\def\D{\displaystyle}				

\def\hbar{h\hspace{-2mm}^-}
\def\kbar{{\mathchar'26\mkern-9muk}}

\def\eps{\varepsilon}
\def\phi{\varphi}
                            
\def\CHI{\mbox{\large $\chi$}}
\def\shalf{\mbox{\small$\frac{1}{2}$\normalsize}}
\def\half{\frac{1}{2}} 
\def\wave{\raisebox{-0.6ex}{\symbol{126}}}

\def\implies{~~~\Rightarrow~~~}
\def\<{\langle} 				
\def\>{\rangle} 				
\def\rangu{\hbox{\raisebox{0.15em}{\rlap{$\sqcap$}}\kern0em%
            \raisebox{-0.19em}{$\sqcup$}} } 

\def\beq{\begin{equation}} 
\def\eeq{\end{equation}} 
\def\lbeq#1{\begin{equation} \label{#1}} 
\def\beqar{\begin{eqnarray}}
\def\eeqar{\end{eqnarray}}
\def\bary{\begin{array}}
\def\eary{\end{array}}
\def\becas{\left\{ \begin{array}{l@{\qquad}l}}
\def\ecas{\end{array} \right.}
\def\benu{\begin{enumerate}}
\def\eenu{\end{enumerate}}
\def\gzit#1{{\rm (\ref{#1})}} 			
\def\fns#1{\mbox{\rm \scriptsize#1}} 		

\def\Cz{\mathbb{C}}

\def\Ez{\mathbb{E}}

\def\Hz{\mathbb{H}}

\def\Rz{\mathbb{R}}

\def\x {{\bf x}}

\begin{document}

\vspace*{-2.5cm}

\begin{center}

{\LARGE \bf Noncommutative analysis and quantum physics} \\
{\LARGE \bf I. Quantities, ensembles and states}

\vspace{0.5cm}

\centerline{\sl {\large \bf Arnold Neumaier}}


\centerline{\sl Institut f\"ur Mathematik, Universit\"at Wien}
\centerline{\sl Strudlhofgasse 4, A-1090 Wien, Austria}
\centerline{\sl email: neum@cma.univie.ac.at}
\centerline{\sl WWW: http://solon.cma.univie.ac.at/\wave neum/}

\end{center}


{\footnotesize
{\bf Abstract.} 
In this sequence of papers, noncommutative analysis is used to give a 
consistent axiomatic approach to a unified conceptual foundation of 
classical and quantum physics, free of undefined terms.
   \newl
The present Part I defines the concepts of quantities, ensembles, and 
states, clarifies the logical relations and operations for them, 
and shows how they give rise to probabilities and dynamics.
The stochastic and the deterministic features of quantum physics 
are separated in a clear way by consistently distinguishing  between
ensembles (representing stochastic elements) and states (representing 
realistic elements).
   \newl
Ensembles are defined by extending the `probability via expectation' 
approach of Whittle to noncommuting quantities. This approach 
carries no connotations of unlimited repeatability; hence it can be 
applied to unique systems such as the universe.
Precise concepts and traditional results about complementarity, 
uncertainty and nonlocality follow with a minimum of technicalities.
Probabilities are introduced in a generality supporting so-called
effects (i.e., fuzzy events).
   \newl
States are defined as partial mappings that provide reference values 
for certain quantities. An analysis of sharpness properties yields 
well-known no-go theorems for hidden variables. 
By dropping the sharpness requirement, hidden variable theories such 
as Bohmian mechanics can be accommodated, but so-called ensemble states 
turn out to be a more natural realization of a realistic state concept.
The weak law of large numbers explains the emergence of classical 
properties for macroscopic systems.
   \newl
Dynamics is introduced via a one-parameter group of automorphisms.
A detailed conceptual analysis of the dynamics in terms of Poisson 
algebras will follow in the second part of this series.
   \newl
The paper realizes a strong formal implementation of Bohr's 
correspondence principle. In all instances, classical and quantum 
concepts are fully parallel: a single common theory has a classical 
realization and a quantum realization.

\vfill
\begin{flushleft}
{\bf Keywords}: 
axiomatization of physics,
Bell inequality,
Bohmian mechanics, 
complementarity,
correspondence principle, 
deterministic, 
effect,
elements of physical reality, 
ensemble, 
event, 
expectation, 
flow of truth, 
foundations of quantum mechanics, 
Heisenberg picture, 
hidden variables, 
ideal measurement, 
nonlocality,
foundations of probability,
preparation of states,
quantities, 
quantum correlations,
quantum logic, 
quantum probability,
reference value,
Schr\"odinger picture,
sharpness 
spin, 
state, 
state of the universe, 
uncertainty relation, 
weak law of large numbers,
Young measure
\end{flushleft}

{\bf E-print Archive No.}: quant-ph/0001096
   \newl
{\bf 1998\hspace{.4em} PACS Classification}: 03.65.Bz, 05.30.Ch
   \newl
{\bf 2000\hspace{.4em} MSC Classification}: primary 81P10, 
secondary 81S05

}

\newpage 
\section{Introduction} \label{intro}

\hfill\parbox[t]{8.8cm}{\footnotesize

{\em ``Look,'' they say, ``here is something new!''
But no, it has all happened before, long before we were born.}

Kohelet, ca. 250 B.C. \cite{Koh2}

\bigskip
{\em Do not imagine, any more than I can bring myself to imagine, 
that I should be right in undertaking so great and difficult a task.  
Remembering what I said at first about probability, I will do my best 
to give as probable an explanation as any other -- or rather, more 
probable; and I will first go back to the beginning and try to speak 
of each thing and of all.}

Plato, ca. 367 B.C. \cite{Pla} 
}\nopagebreak

\bigskip
This paper is the first one of a series of papers designed to 
give a mathematically elementary and philosophically consistent 
axiomatic foundation of modern theoretical physics, free of undefined 
terms. It is an attempt to reconsider, from the point of view of 
noncommutative analysis, {\sc Hilbert}'s \cite{Hil} sixth problem, 
the {\em axiomatization of theoretical physics}. 
(It is an attempt only since at the present stage 
of development, I have not yet tried to achieve full mathematical 
rigor everywhere. However, the present Part I is completely rigorous,
and in later parts the few places where the standard of rigor is 
relaxed will be explicitly mentioned.)

The purpose is to provide precise mathematical concepts that match all
concepts that physicists use to describe their experiments and their
theory, in sufficiently close correspondence to reproduce at least 
that part of physics that is amenable to numerical verification.

One of the basic premises of this work is that the split between 
classical physics and quantum physics should be as small as possible. 
Except in the examples, the formalism never distinguishes between the 
classical and the quantum situation. Thus it can be considered as a 
consequent implementation of {\sc Bohr}'s {\em correspondence 
principle}. This also has didactical advantages for teaching: 
Students can be trained to be acquainted with the formalism by means 
of intuitive, primarily classical examples at first. Later, without 
having to unlearn anything, they can apply the same formalism to 
quantum phenomena.

\bigskip
The present Part I is concerned with giving (more carefully than usual, 
and without reference to measurement) a concise foundation by 
defining the concepts of quantities, ensembles, and states, 
clarifying the logical relations and operations for them, and showing 
how they give rise to the traditional postulates of quantum mechanics, 
including probabilities and dynamics.

The stochastic and the deterministic features of quantum physics 
are separated in a clear way by consistently distinguishing  between
ensembles (representing stochastic elements) and states (representing 
realistic elements).

Most of what is done here is common wisdom in quantum mechanics; 
see, e.g., {\sc Jammer} \cite{Jam1,Jam2}, {\sc Jauch} \cite{Jau}, 
{\sc Messiah} \cite{Mes}, {\sc von Neumann} \cite{vNeu}.

However, the new interpretation slightly shifts the meaning of the
concept of a state, fixing it in a way that allows to embed and
analyze different interpretations of the quantum mechanical formalism,
including both orthodox views such as the Copenhagen interpretation 
and hidden-variable theories such as Bohmian mechanics.

To motivate the conceptual foundation and to place it into context,
I found it useful to embed the formalism into my philosophy of physics,
while {\em strictly separating the mathematics by using a formal
definition-example-theorem-proof exposition style}. Though I present my 
view generally without using subjunctive formulations or qualifying 
phrases, I do not claim that this is the only way to understand physics.
However, I did attempt to integrate different points of view. And I 
believe that my philosophical view is consistent with the mathematical 
formalism of quantum mechanics and accommodates naturally a number of
puzzling questions about the nature of the world.

\bigskip
The stochastic contents of quantum theory is determined by the 
restrictions noncommutativity places upon the preparation of 
experiments. Since the information going into the preparation is 
always extrapolated from finitely many observations in the past, 
it can only be described in a statistical way, i.e., by ensembles. 

Ensembles are defined by extending to noncommuting quantities
{\sc Whittle}'s \cite{Whi} elegant expectation approach to classical
probability theory. This approach carries no connotations of unlimited 
repeatability; hence it can be applied to unique systems such as the 
universe. The weak law of large numbers relates abstract ensembles and 
concrete mean values over many instances of quantities with the same 
stochastic behavior within a single system.

Precise concepts and traditional results about complementarity, 
uncertainty and nonlocality follow with a minimum of technicalities. 
In particular, nonlocal correlations predicted by {\sc Bell} \cite{Bel} 
and first detected by {\sc Aspect} \cite{Asp} are shown to be already 
consequences of the nature of quantum mechanical ensembles and do not 
depend on hidden variables or on counterfactual reasoning.

The concept of probability itself is derived from that of an ensemble 
by means of a formula motivated from classical ensembles that can be 
described as a finite weighted mean of properties of finitely many 
elementary events.
Probabilities are introduced in a generality supporting so-called
effects, a sort of fuzzy events (related to POV measures that play a 
significant role in measurement theory; see {\sc Busch} et al. 
\cite{BusGL,BusLM}, {\sc Davies} \cite{Dav}, {\sc Peres} \cite{Per3}). 
The weak law of large numbers provides
the relation to the frequency interpretation of probability.
As a special case of the definition, one gets without any effort the
well-known squared probability amplitude formula for transition 
probabilities.

\bigskip
States are defined as partial mappings that provide objective 
reference values for certain quantities. Sharpness of quantities is
defined in terms of laws for the reference values, in particular
the squaring law that requires the value of a squared sharp quantity 
$f$ to be equal to the squared value of $f$. It is shown that
the values of sharp quantities must belong to their spectrum, and
that requiring all quantities to be sharp produces contradictions
over Hilbert spaces of dimension $>3$. This is related to well-known 
no-go theorems for hidden variables. (However, recent constructive 
results by {\sc Clifton \& Kent} \cite{CliK} show that in the 
finite-dimensional case there are states with a dense set of sharp 
quantities.)

An analysis of a well-known macroscopic reference value, the center
of mass, leads us to reject the sharpness requirement. Without
universal sharpness, hidden variable theories such as Bohmian 
mechanics ({\sc Bohm} \cite{Boh}; cf. {\sc Holland} \cite{Hol}) can be 
accommodated. However, the Bohmian states violate monotony, and 
so-called ensemble states turn out to be a more natural realization of 
a realistic state concept. 

With ensemble states, quantum objects are intrinsically extended, real 
objects; e.g., the reference radius of a hydrogen atom in the ground 
state is 1.5 times the Bohr radius. Moreover, in ensemble states, 
the weak law of large numbers explains the emergence of classical 
properties for macroscopic systems.

Thus ensemble states provide an elegant solution to the reality 
problem, confirming the insistence of the orthodox Copenhagen 
interpretation on that there is nothing but ensembles, while avoiding 
their elusive reality picture. 

\bigskip
Finally, it is outlined how dynamical properties fit into the present
setting. Dynamics is introduced via a one-parameter group of 
automorphisms. A detailed conceptual analysis of the dynamics in terms 
of a differential calculus based on Poisson algebras will follow in the second part of this series.
 
Subsequent parts of this sequence of papers will present 
the calculus of integration and its application to 
equilibrium thermodynamics, a theory of measurement,
a relativistic covariant Hamiltonian multiparticle theory, 
and its application to nonequilibrium thermodynamics 
and field theory.

As in this first paper, each topic will be presented in a uniform way, 
classical and quantum versions being only special cases of a single 
theory.

\bigskip
{\bf Acknowledgments.} 
I'd like to thank Waltraud Huyer, Willem de Muynck, Hermann Schichl, 
Tapio Schneider, Victor Stenger, Karl Svozil and Roderich Tumulka 
for useful discussions of an earlier version of this manuscript.

\section{Quantities} \label{quantities}

\hfill\parbox[t]{8.8cm}{\footnotesize

{\em Only love transcends our limitations. In contrast, our predictions
can fail, our communication can fail, and our know\-ledge can fail.
For our knowledge is patchwork, and our predictive power is 
limited. But when perfection comes, all patchwork will disappear.}

St. Paul, ca. 57 A.D. \cite{Pau}

\bigskip
{\em But you} [God] 
{\em have ordered everything with measure, number and weight.} 
             
Wisdom 11:20, ca. 50 B.C.
}\nopagebreak

\bigskip

All our scientific knowledge is based on past observation, and only 
gives rise to conjectures about the future. Mathematical consistency 
requires that our choices are constrained by some formal laws. When we 
want to predict something, the true answer depends on knowledge we do 
not have. We can calculate at best approximations whose accuracy 
can be estimated using statistical techniques (assuming that the 
quality of our models is good).

This implies that we must distinguish between {\em quantities} 
(formal concepts of what can possibly be measured or calculated) and 
{\em numbers} (the results of measurements and calculations 
themselves); those quantities that are constant by the nature of the 
concept considered behave just like numbers. 

Physicists are used to calculating with quantities that they may add 
and multiply without restrictions; if the quantities are complex, the 
complex conjugate can also be formed. It must also be possible to 
compare quantities, at least in certain cases.

Therefore we take as primitive objects of our treatment a set $\Ez$ of 
quantities, such that the sum and the product of quantities is again a 
quantity, and there is an operation generalizing complex conjugation. 
Moreover, we assume that there is an ordering relation that allows us 
to compare two quantities.

Operations on quantities and their comparison are required to satisfy 
a few simple rules; they are called {\bf axioms} since we take them as 
a formal starting point without making any further demands on the
nature of the symbols we are using. Our axioms are motivated by the 
wish to be as general as possible while still preserving the ability 
to manipulate quantities in the manner familiar from matrix algebra. 
(Similar axioms for quantities have been proposed, e.g.,
by {\sc Dirac} \cite{Dir}.)

\begin{dfn} ~

\nopagebreak
(i) $\Ez$ denotes a set whose elements are called {\bf quantities}.
For any two quantities $f,g\in\Ez$, the {\bf sum} $f+g$, the 
{\bf product} $fg$, and the {\bf conjugate} $f^*$ are also quantities.
It is also specified for which pairs of quantities the relation 
$f\geq g$ holds.

The following axioms (Q1)--(Q8) are assumed to hold for all complex 
numbers $\alpha\in\Cz$ and all quantities $f,g,h\in\Ez$.

(Q1) 
~$\Cz \subseteq \Ez$, i.e., complex numbers are special quantities,
where addition, multiplication and conjugation have their traditional
meaning. 

(Q2)
~{$(fg)h=f(gh)$,~~ $\alpha f=f\alpha $,~~ $0f=0$,~~ $1f=f$.}

(Q3)
~{$(f+g)+h=f+(g+h)$,~~ $f(g+h)=fg+fh$,~~ $f+0=f$.}

(Q4)
~{$f^{**}=f$,~~ $(fg)^* =g^* f^* $,~~ $(f+g)^* =f^* +g^*$.}

(Q5)
~{$f^* f =0 \implies f =0$.}

(Q6) 
~$\geq$ is a partial order, i.e., it is reflexive ($f\geq f$),
antisymmetric ($f\geq g \geq f \Rightarrow f=g$) and transitive 
($f\geq g \geq h \Rightarrow f \geq h)$.

(Q7)
~{$f\geq g \implies f+h\geq g+h$.}

(Q8)
~{$f\geq 0 \implies f=f^*$ and $g^*fg\geq 0$.}

(Q9)
~ $1 \geq 0$.

If (Q1)--(Q9) are satisfied we say that $\Ez$ is a {\bf Q-algebra}.

(ii) We introduce the traditional notation
\[ 
f \leq g :\Leftrightarrow g\geq f,
\]
\[
-f:=(-1)f,~~ f-g:=f+(-g), ~~~[f,g]:=fg-gf,
\]
\[
f^0:=1,~~ f^l:=f^{l-1}f~~~ (l=1,2,\dots ),
\]
\[
\re f = \half(f+f^*),~~~\im f = \frac{1}{2i}(f-f^*),
\]
\[
\|f\|=\inf\{\alpha\in\Rz \mid f^*f \leq \alpha^2, \alpha\geq0 \}.
\]
(The infimum of the empty set is taken to be $\infty$.)
$[f,g]$ is called the {\bf commutator} of $f$ and $g$, $\re f$, 
$\im f$ and $\|f\|$ are referred to as the {\bf real part}, the 
{\bf imaginary part}, and the {\bf (spectral) norm} of $f$, 
respectively. The {\bf uniform topology} is the topology induced on
$\Ez$ by declaring a set $E$ open if it contains a ball
$\{f\in\Ez \mid \|f\|<\eps\}$ for some $\eps>0$.

(iii) A quantity $f\in\Ez$ is called {\bf bounded} if 
$\|f\|<\infty$, 
{\bf Hermitian} if $f^*=f$,
and {\bf normal} if $[f,f^*]=0$. More generally, a set $F$ of 
quantities is called {\bf normal} if all its quantities commute with 
each other and with their conjugates.
\end{dfn}

Note that every Hermitian quantity (and in a commutative algebra, 
every quantity) is normal. Physical observables will be among the 
normal quantities, but until we define (in a later part of this 
sequence of papers) what it means to `observe' a quantity we avoid 
talking about observables.

\begin{expls}~

(i) The commutative algebra $\Ez = \Cz^n$ with pointwise multiplication
and componentwise inequalities is a Q-algebra, if vectors with 
constant entries $\alpha$ are identified with $\alpha\in\Cz$.
This Q-algebra describes properties of $n$ classical elementary events;
cf. Example \ref{ex.5.3}(i).

(ii) $\Ez=\Cz^{n\times n}$ is a Q-algebra if complex numbers are 
identified with the scalar multiples of the identity matrix, and
$f\geq g$ iff $f-g$ is Hermitian and positive semidefinite.
This Q-algebra describes quantum systems with $n$ levels. 
For $n=2$, it also describes a single spin, or a qubit.

(iii) The algebra of all complex-valued functions on a set $\Omega$,
with pointwise multiplication and pointwise inequalities is a 
Q-algebra. Suitable subalgebras of such algebras describe classical
probability theory -- cf. Example \ref{ex1.4}(i) -- and classical 
mechanics -- cf. Example \ref{ex.classquant}(i). In the latter case,
$\Omega$ is the phase space of the system considered.

(iv) The algebra of bounded linear operators on a Hilbert space 
$\Hz$, with $f\geq g$ iff $f-g$ is Hermitian and positive semidefinite,
is a Q-algebra. They (or the more general $C^*$-algebras and von 
Neumann algebras) are frequently taken as the basis of nonrelativistic 
quantum mechanics.

(v) The algebra of continuous linear operators on the Schwartz space 
${\cal S}(\Omega_{qu})$ of rapidly decaying functions on a manifold 
$\Omega_{qu}$ is a Q-algebra. It also allows the discussion of 
unbounded quantities. In quantum physics, $\Omega_{qu}$ is the 
configuration space of the system.

Note that physicist generally need to work with unbounded quantities, 
while much of the discussion on foundations takes the more restricted 
Hilbert space point of view. The theory presented here is formulated 
in a way to take care of unbounded quantities, while in our examples, 
we select the point of view as deemed profitable.

\end{expls}

We shall see that, for the general, qualitative aspects of the theory
there is no need to know any details of how to actually perform 
calculations with quantities; this is only needed if one wants to 
calculate specific properties for specific systems. In this respect, 
the situation is quite similar to the traditional axiomatic treatment of
real numbers: The axioms specify the permitted ways to handle formulas 
involving these numbers; and this is enough to derive calculus, say,
without the need to specify either what real numbers {\em are} or 
algorithmic rules for addition, multiplication and division. Of course,
the latter are needed when one wants to do specific calculations but not
while one tries to get insight into a problem. And as the development
of pocket calculators has shown, the capacity for understanding theory 
and that for knowing the best ways of calculation need not even reside 
in the same person.

Note that we assume commutativity only between numbers and quantities.
However, general commutativity of the addition is a consequence of our 
other assumptions. We prove this together with some other useful 
relations. 

\begin{prop}
For all quantities  $f$, $g$, $h\in \Ez$ and $\lambda \in\Cz$,
\lbeq{e.p1}
(f+g)h=fh+gh,~~f-f=0,~~ f+g=g+f
\eeq
\lbeq{e.p2}
[f,f^*]=-2i[\re f,\im f],
\eeq
\lbeq{e.p3}
f^*f\geq 0,~~ ff^*\geq 0.
\eeq
\lbeq{e.p4}
f^*f\leq 0 \implies \|f\|=0 \implies f=0,
\eeq
\lbeq{e.p5}
f\leq g \implies h^*fh\leq h^*gh,~|\lambda|f\leq|\lambda|g,
\eeq
\lbeq{e.p6}
f^*g+g^*f\leq 2\|f\|~\|g\|,
\eeq
\lbeq{e.p7}
\|\lambda f\|=|\lambda| \|f\|,~~~ \|f\pm g\|\leq \|f\|\pm \|g\|,
\eeq
\lbeq{e.p8}
\|f g\|\leq \|f\|~ \|g\|.
\eeq
\end{prop}

\bepf
The right distributive law follows from
\[
\begin{array}{lll}
(f+g)h&=&((f+g)h)^{* *}=(h^* (f+g)^* )^* =(h^* (f^* +g^* ))^* \\
&=&(h^* f^* +h^* g^* )^* =(h^* f^* )^* +(h^* g^* )^* \\
&=&f^{* * }h^{* * }+g^{* * }h^{* * }=fh+gh.
\end{array}
\]
It implies $f-f=1f-1f=(1-1)f=0f=0$. From this, we may deduce that 
addition is commutative, as follows. The quantity $h:=-f+g$
satisfies
\[
-h=(-1)((-1)f+g)=(-1)(-1)f+(-1)g=f-g, 
\]
and we have
\[
f+g=f+(h-h)+g=(f+h)+(-h+g)=(f-f+g)+(f-g+g)=g+f. 
\]
This proves \gzit{e.p1}. If $u=\re f$, $v=\im f$ then $u^*=u,v^*=v$
and $f=u+iv, f^*=u-iv$. Hence 
\[
[f,f^*]=(u+iv)(u-iv)-(u-iv)(u+iv)=2i(vu-uv)=-2i[\re f,\im f],
\]
giving \gzit{e.p2}. \gzit{e.p3}--\gzit{e.p5} follow directly from 
(Q7) -- (Q9). Now let $\alpha=\|f\|$, $\beta=\|g\|$. Then 
$f^*f\leq \alpha^2$ and $g^*g\leq \beta^2$. Since
\[
\begin{array}{lll}
0\leq (\beta f - \alpha g)^*(\beta f - \alpha g)&=&
\beta^2f^*f-\alpha\beta(f^*g+g^*f)+\alpha^2 g^*g\\
&\leq& \beta^2\alpha^2 \pm\alpha\beta(f^*g+g^*f) +\alpha^2 g^*g,
\end{array}
\] 
$f^*g+g^*f\leq 2\alpha\beta$ if $\alpha\beta\neq 0$, and for
$\alpha\beta=0$, the same follows from \gzit{e.p4}. Therefore
\gzit{e.p6} holds. The first half of \gzit{e.p7} is trivial, and
the second half follows for the plus sign from 
\[
(f+g)^*(f+g)=f^*f+f^*g+g^*f+g^*g
\leq \alpha^2+ 2\alpha\beta+\beta^2=(\alpha+\beta)^2,
\]
and then for the minus sign from the first half.
Finally, by \gzit{e.p5},
\[
(fg)^*(fg)=g^*f^*fg\leq g^*\alpha^2g=\alpha^2g^*g\leq\alpha^2\beta^2.
\]
This implies \gzit{e.p8}.
\epf

\begin{cor}\label{c1.3}~

(i) Among the complex numbers, precisely the nonnegative real numbers
$\lambda$ satisfy $\lambda\geq 0$.

(ii) For all $f\in\Ez$, $\re f$ and $\im f$ are Hermitian. $f$ is 
Hermitian iff $f=\re f$ iff $\im f=0$. If $f,g$ are commuting 
Hermitian quantities then $fg$ is Hermitian, too.

(iii) $f$ is normal iff $[\re f,\im f]=0$.
\end{cor}

\bepf
(i) If $\lambda$ is a nonnegative real number then $\lambda=f^*f\geq0$ 
with $f=\sqrt{\lambda}$. If $\lambda$ is a negative real number then 
$\lambda=-f^*f\leq0$ with $f=\sqrt{-\lambda}$, and by antisymmetry,
$\lambda\geq0$ is impossible. If $\lambda$ is a nonreal number then 
$\lambda\neq\lambda^*$ and $\lambda\geq0$ is impossible by (Q8).

The first two assertions of (ii) are trivial, and the third holds since
$(fg)^*=g^*f^*=gf=fg$ if $f,g$ are Hermitian and commute.

(iii) follows from \gzit{e.p2}. 
\epf

Thus, in conventional terminology (see, e.g., {\sc Rickart} \cite{Ric}),
$\Ez$ is a {\bf partially ordered nondegenerate *-algebra with unity},
but not necessarily with a commutative multiplication. 

\begin{rem} 
In the realizations of the axioms I know of, e.g., in $C^*$-algebras 
({\sc Rickart} \cite{Ric}), we also have the relations
\[
\|f^*\|=\|f\|,~~~\|f^*f\|=\|f\|^2,
\]
and
\[
0\leq f \leq g \implies f^2 \leq g^2,
\]
but I have not been able to prove these from the present axioms,
and they were not needed to develop the theory.
\end{rem}

As the example
$\Ez=\Cz^{n\times n}$ shows, $\Ez$ may have zero
divisors, and not every nonzero quantity need have an inverse.
Therefore, in the manipulation of formulas, precisely the same 
precautions must be taken as in ordinary matrix algebra.

\section{Complementarity} \label{compl}

\hfill\parbox[t]{8.8cm}{\footnotesize

{\em You cannot have the penny and the cake.}

Proverb

}\nopagebreak

\bigskip
The lack of commutativity gives rise to the phenomenon of
complementarity, expressed by inequalities that demonstrate the danger 
of simply thinking of quantities in terms of numbers. 

\begin{dfn}
Two Hermitian quantities $f,g$ are called {\bf complementary} if there
is a real number $\gamma>0$ such that
\lbeq{e.compl}
(f-x)^2+(g-y)^2 \geq \gamma^2~~~\mbox{for all }x,y\in\Rz.
\eeq
\end{dfn}

\begin{expls}~

(i) The Q-algebra of all complex-valued functions on a set $\Omega$
contains no complementary pair of quantities. Indeed, setting
$x=f(\omega)$, $y=g(\omega)$ in \gzit{e.compl}, we find 
$0\geq \gamma^2$, contradicting complementarity.

Thus complementarity captures the phenomenon where two quantities do 
not have simultaneous sharp classical `values'. 
(See also Section \ref{states}.)

(ii) $\Cz^{2\times2}$ contains a complementary pair of quantities.
Indeed, the Pauli matrices
\lbeq{e.pauli}
\sigma_1 =\left(\begin{array}{l}0~~1\\1~~0\end{array}\right),~~
\sigma_3 =\left(\begin{array}{l}1~~\phantom{-}0\\0~~-1
\end{array}\right)
\eeq
are complementary; see Proposition \ref{p.comp}(i) below.

(iii) The algebra of bounded linear operators on a Hilbert space of 
dimension $>1$ contains a complementary pair of quantities, since
it contains many subalgebras isomorphic to $\Cz^{2\times2}$.

(iv) In the algebra of all linear operators on the Schwartz space 
${\cal S}(\Rz)$, {\bf position} $q$, defined by 
\[
(qf)(x)=xf(x),
\]
and {\bf momentum} $p$, defined by 
\[
(pf)(x)=-i\hbar f'(x),
\]
where $\hbar>0$ is Planck's constant, are complementary.
Since $q$ and $p$ are Hermitian, this follows from the easily
verified {\bf canonical commutation relation}
\lbeq{ccr}
[q,p]=i\hbar
\eeq
and Proposition \ref{p.comp}(ii) below.

\end{expls}

\begin{prop}\label{p.comp}~

(i) The Pauli matrices \gzit{e.pauli} satisfy
\lbeq{e6.uncpauli}
(\sigma_1-s_1)^2+(\sigma_3-s_3)^2 \geq 1 
~~~\mbox{for all } s_1,s_3\in\Rz.
\eeq

(ii) Let $p,q$ be Hermitian quantities satisfying $[q,p]=i\hbar$.
Then, for any $k,x\in\Rz$ and any positive $\Delta p,\Delta q \in\Rz$,
\lbeq{e6.unc}
\Big(\frac{p-k}{\Delta p}\Big)^2+\Big(\frac{q-x}{\Delta q}\Big)^2
\geq \frac{\hbar}{\Delta p \Delta q}.
\eeq
\end{prop}
\begin{proof}
(i) A simple calculation gives
\[
(\sigma_1-s_1)^2+(\sigma_3-s_3)^2-1=\left(\begin{array}{cc}
s_1^2+(1-s_3)^2 & -2s_1 \\
-2s_1           & s_1^2+(1+s_3)^2 \\
\end{array}\right) \geq 0,
\]
since the diagonal is nonnegative and the determinant is
$(s_1^2+s_3^2-1)^2\geq 0$.

(ii) The quantities $f=(q-x)/\Delta q$ and $g=(p-k)/\Delta p$ are 
Hermitian and satisfy $[f,g]=[q,p]/\Delta q\Delta p=i\kappa$ where 
$\kappa=\hbar/\Delta q\Delta p$. Now \gzit{e6.unc} follows from
\[
0\leq (f+ig)^*(f+ig)=f^2+g^2+i[f,g]=f^2+g^2-\kappa.
\]
\end{proof}

The complementarity of position and momentum expressed by \gzit{e6.unc0}
is the deeper reason for the Heisenberg uncertainty relation discussed 
later in \gzit{e6.unc0} and \gzit{e6.unc1}.

\begin{thm}
In $\Cz^{n\times n}$, two complementary quantities cannot commute.
\end{thm}

\bepf
Any two commuting quantities $f,g$ have a common eigenvector $\psi$.
If $f\psi=x\psi$ and $g\psi=y\psi$ then $\psi^*((f-x)^2+(g-y)^2)\psi=0$, 
whereas \gzit{e.compl} implies 
\[
\psi^*(f-x)^2+(g-y)^2)\psi\geq \gamma^2\psi^*\psi >0.
\] 
Thus $f,g$ cannot be complementary.
\epf

I have not been able to decide whether complementary quantities can 
possibly commute. (It is impossible when there is a joint spectral 
resolution.)

\section{Ensembles} 
\label{ensembles}

\hfill\parbox[t]{8.8cm}{\footnotesize

{\em We may assume that words are akin to the matter which they 
describe; when they relate to the lasting and permanent and 
intelligible, they ought to be lasting and unalterable, and, as far 
as their nature allows, irrefutable and immovable -- nothing less.  
But when they express only the copy or likeness and not the eternal 
things themselves, they need only be likely and analogous to the real 
words. As being is to becoming, so is truth to belief.}

Plato, ca. 367 B.C. \cite{Pla}
}\nopagebreak

\bigskip
The stochastic nature of quantum mechanics is usually discussed in
terms of {\em probabilities}. However, from a strictly logical point 
of view, this has the drawback that one gets into conflict with the 
traditional foundation of probability theory by 
{\sc Kolmogorov} \cite{Kol}, which does not extend to the 
noncommutative case. Mathematical physicists (see, e.g., 
{\sc Parthasarathy} \cite{Par}, {\sc Meyer} \cite{Mey}) developed a 
far reaching quantum probability calculus based on Hilbert space 
theory. But their approach is highly formal, drawing its motivation 
from analogies to the classical case rather than from the common 
operational meaning.

{\sc Whittle} \cite{Whi} presents a much less known alternative
approach to classical probability theory, equivalent to that 
of Kolmogorov, that treats {\em expectation} as the basic concept and
derives probability from axioms for the expectation. (See the 
discussion in \cite[Section 3.4]{Whi} why, for historical reasons, 
this has remained a minority approach.)  

The approach via expectations is easy to motivate, leads quickly to 
interesting results, and extends without much trouble to the quantum 
world, yielding the ensembles (`mixed states') of traditional quantum 
physics. As we shall see, explicit probabilities enter only at 
a very late stage of the development.

A significant advantage of the expectation approach compared with the 
probability approach is that it is intuitively more removed from 
connotations of `unlimited repeatability'. Hence it can be naturally 
used for {\em unique} systems such as the set of all natural 
globular proteins (cf., e.g., {\sc Neumaier} \cite{Neu.prot}), the 
climate of the earth, or the universe, and to deterministic, 
pseudo-random behavior such as rounding errors in floating point 
computations (cf., e.g., {\sc Higham} \cite[Section 2.6]{Hig}), once 
these have enough complexity to exhibit finite {\em internal} 
repetitivity to which the weak law of large numbers 
(Theorem \ref{t.weaklaw} below) may be applied.

\bigskip
The axioms we shall require for meaningful expectations are those
trivially satisfied for weighted averages of a finite ensemble of 
observations. While this motivates the form of the axioms and the
name `ensemble' attached to the concept, there is no need at all to 
interpret expectation as an average; this is the case only in 
certain classical situations. In general, ensembles are simply a way to 
consistently organize structured data obtained by some process of 
observation. 

For the purpose of statistical analysis and prediction, it is 
completely irrelevant what this process of observation entails. 
What matters is only that for certain quantities observed values are 
available that can be compared with their expectations. 
The expectation of a quantity $f$ is simply a value near which, based 
on the theory, we should expect an observed value for $f$. At the same 
time, the standard deviation serves as a measure of the amount to 
which we should expect this nearness to deviate from exactness. 

For science, however, it is of utmost importance to have well-defined 
protocols that specify what are valid observations. Such standardized 
protocols guarantee that the observations are repeatable and hence 
objective. On the other hand, these protocols require a level of 
description not appropriate for the foundations of a discipline.
Therefore, at the present fundamental level of exposition, observed 
values are undefined, and not yet part of the formal development. 
In physics, they need a theory of measurement, which will be discussed
in a later part of this sequence of papers.

\begin{dfn}~

\nopagebreak
(i) An {\bf ensemble} is a mapping $^-$ that assigns to each quantity 
$f \in \Ez$ its {\bf expectation} $\overline{f}=:\< f\> \in \Cz$ 
such that for all $f,g \in \Ez$, $\alpha \in \Cz$,

(E1)~ $\<1\> =1, ~~\<f^*\>=\<f\>^*,~~ \< f+g\> =\<f\> +\<g\> $, 

(E2)~ $\<\alpha f\> =\alpha\<f\>$, 

(E3)~ If $f \ge 0$ then $\<f\> \ge 0$,

(E4)~ If $f_l\in\Ez,~ f_l \downarrow 0$ then $\inf \<f_l\> = 0$.

Here $f_l \downarrow 0$ means that the $f_l$ converge almost
everywhere to $0$ and $f_{l+1}\leq f_l$ for all $l$.

(ii) The number
\[
\cov(f,g):=\re \<(f-\overline{f})^*(g-\overline{g}) \>
\]
is called the {\bf covariance} of $f,g\in\Ez$. Two quantities $f,g$ are 
called {\bf correlated} if $\cov(f,g)\neq0$, and {\bf uncorrelated} 
otherwise.

(iii) The number 
\[
\sigma(f):=\sqrt{\cov(f,f)}
\]
is called the {\bf uncertainty} or {\bf standard deviation} of 
$f\in\Ez$ in the ensemble $\<\cdot\>$. 

\end{dfn}

(We shall not use axiom (E4) in this paper and therefore defer
technicalities about almost everywhere convergence to
a more detailed treatment in a later part of this sequence of papers).

This definition generalizes the expectation axioms of
{\sc Whittle} \cite[Section 2.2]{Whi} for classical probability theory
and the definitions of elementary classical statistics.
Note that (E3) ensures that $\sigma(f)$ is a nonnegative real number
that vanishes if $f$ is a constant quantity (i.e., a complex number).

\begin{expls}  \label{ex.5.3}~

(i) {\bf Finite probability theory.}
In the commutative Q-algebra $\Ez = \Cz^n$ with pointwise 
multiplication and componentwise inequalities, every linear functional 
on $\Ez$, and in particular every ensemble, has the form 
\lbeq{e5.fin}
\<f\>=\sum_{k=1}^n p_k f_k
\eeq
for certain weights $p_k$. The ensemble axioms hold precisely 
when the $p_k$ are nonnegative and add up to one; thus $\<f\>$ 
is a weighted average, and the weights have the intuitive meaning of 
`probabilities'. 

Note that the weights can be recovered from the expectation
by means of the formula $p_k=\<e_k\>$, where $e_k$ is the unit vector
with a one in component $k$.

(ii) {\bf Quantum mechanical ensembles.}
In the Q-algebra $\Ez$ of bounded linear operators on a Hilbert space 
$\Hz$, quantum mechanics describes a {\bf pure ensemble} 
(traditionally called a `pure state', but we shall reserve the name
`state' for a concept defined in Section \ref{states}) 
by the expectation
\[
\<f\>:=\psi^*f\psi,
\]
where $\psi\in\Hz$ is a unit vector. And quantum thermodynamics 
describes an {\bf equilibrium ensemble} by the expectation
\[
\<f\>:=\tr e^{-S/\kbar}f, 
\]
where $\kbar>0$ is the {\bf Boltzmann constant}, and $S$ is a Hermitian
quantity with $\tr e^{-S/\kbar}=1$ called the {\bf entropy} whose 
spectrum is discrete and bounded below.
In both cases, the ensemble axioms are easily verified.

\end{expls}

\begin{prop} \label{p5.2}

For any ensemble,

(i) $f\leq g \implies \<f\> \leq \<g\>$.

(iii) For $f,g\in\Ez$,
\[
\cov(f,g)=\re(\<f^*g\>-\<f\>^*\<g\>),
\]
\[
\<f^*f\>=\<f\>^*\<f\>+\sigma(f)^2,
\]
\[
|\<f\>|\leq\sqrt{\<f^*f\>}.
\]

(iii) If $f$ is Hermitian then $\bar f = \<f\>$ is real and
\[
\sigma(f)=\sqrt{\<(f-\overline{f})^2 \>}
=\sqrt{\<f^2\>-\<f\>^2}.
\]

(iv) Two commuting Hermitian quantities $f,g$ are uncorrelated iff
\[
\<fg\>=\<f\>\<g\>.
\]

\end{prop}

\bepf
(i) follows from (E1) and (E3).

(ii) The first formula holds since
\[
\<(f-\bar f)^*(g-\bar g)\>
=\<f^*g\>-\bar f^*\<g\>-\<f\>^*\bar g +\bar f^*\bar g 
= \<f^*g\>-\<f\>^*\<g\>.
\]
The second formula follows for $g=f$, using (E1), and the third 
formula is an immediate consequence.

(iii) follows from (E1) and (ii).

(iv) If $f,g$ are Hermitian and commute the $fg$ is Hermitian by 
Corollary \ref{c1.3}(ii), hence $\<fg\>$ is real. By (iii),
$\cov(f,g)=\<fg\>-\<f\>\<g\>$, and the assertion follows.

\epf

Fundamental for the practical use of ensembles, and basic to 
statistical mechanics, is the {\bf weak law of large numbers}:

\begin{thm}\label{t.weaklaw} 
For a family of quantities $f_l$ $(l=1, \ldots , N)$ with
constant expectation $\< f_l \> = \mu$, the {\bf mean value}
\[
  \bar f := \frac{1}{N} \D \sum ^N _{l=1} f_l
\]
satisfies
\[
  \< \bar f \> =\mu.
\]
If, in addition, the $f_l$ are uncorrelated and have constant standard 
deviation $\sigma(f_l)=\sigma$ then
\lbeq{e.sigN}
\sigma (\bar f) = \sigma/\sqrt{N}
\eeq
becomes arbitrarily small as $N$ becomes sufficiently large. 
\end{thm}

\bepf 
We have
\[
\<\bar f\> =\frac{1}{N}(\<f_1\>+\dots+\<f_N\> )
=\frac{1}{N}(\mu+\dots+\mu)=\mu
\] 
and
\[
\bar f^*\bar f=\frac{1}{N^2}\Big(\sum_jf_j\Big)^*\Big(\sum_kf_k\Big)
=N^{-2}\sum_{j,k}f_j^*f_k.
\]
Now
\[
\<f_j^*f_j\>=\<f_j\>^*\<f_j\>+\sigma(f_j)^2=|\mu|^2+\sigma^2
\]
and, if the $f_l$ are uncorrelated, for $j\neq k$,
\[
\<f_j^*f_k+f_k^*f_j\>=2\re \<f_j^*f_k\>
=2\re \<f_j\>^*\<f_k\>=2\re \mu^*\mu=2|\mu|^2.
\]
Hence 
\[
\begin{array}{lll}
\sigma(\bar f)^2 &=& \<\bar f^*\bar f\>-\<\bar f\>^*\<\bar f\> \\
&=& N^{-2}\Big(N(\sigma^2+|\mu|^2)+{N \choose 2}2|\mu|^2\Big)-\mu^*\mu
=N^{-1}\sigma^2,
\end{array}
\]
and the assertions follow.
\epf

\section{Uncertainty} 
\label{uncertainty}

\hfill\parbox[t]{8.8cm}{\footnotesize

{\em For you do not know which will succeed, whether this or that, or 
whether both will do equally well.}

Kohelet, ca. 250 B.C. \cite{Koh}

}\nopagebreak

\bigskip
Due to our inability to 
prepare experiments with a sufficient degree of sharpness to know 
with certainty everything about a system we investigate,
we need to describe the preparation of experiments in a 
stochastic language that permits the discussion of such uncertainties; 
in other words, we shall model prepared experiments by ensembles.

Formally, the essential difference between classical mechanics 
and quantum mechanics in the latter's lack of commutativity.
While in classical mechanics there is in principle no lower
limit to the uncertainties with which we can prepare the quantities
in a system of interest,
the quantum mechanical uncertainty relation for noncommuting 
quantities puts strict limits on the uncertainties in the preparation
of microscopic ensembles. Here, {\em preparation} is defined informally 
as bringing the system into an ensemble such that measuring certain 
quantities gives values that agree with the expectation to an accuracy 
specified by given uncertainties.

In this section, we discuss the limits of the accuracy to which this 
can be done.

\begin{prop} ~\nopagebreak

(i) The {\bf Cauchy--Schwarz inequality}  
\[
|\< f^*g \>|^2 \le \< f^*f \>\< g^*g \>
\]
holds for all $f,g\in\Ez$.

(ii) The {\bf uncertainty relation}
\[
\sigma(f)^2\sigma(g)^2 
\geq |\cov(f,g)|^2+\left|\shalf\<f^*g-g^*f\>\right|^2
\]
holds for all $f,g\in\Ez$.

(iii) For $f,g\in\Ez$, 
\lbeq{ecov1}
\cov(f,g)=\cov(g,f)=\shalf(\sigma(f+g)^2-\sigma(f)^2-\sigma(g)^2),
\eeq
\lbeq{ecov}
|\cov(f,g)| \leq \sigma(f)\sigma(g), 
\eeq
\lbeq{esig}
\sigma(f+g) \leq \sigma(f)+\sigma(g).
\eeq
In particular,
\lbeq{e.prodbound}
|\<fg\>-\<f\>\<g\>|\leq\sigma(f)\sigma(g) 
~~~\mbox{for commuting Hermitian } f,g. 
\eeq

\end{prop}

\bepf
(i) For arbitrary $\alpha ,\beta\in \Cz$ we have
\[
\begin{array}{ll}
0&\le \<(\alpha f-\beta g)^*(\alpha f-\beta g )\> \\
&=\alpha ^* \alpha \< f^*f \>-\alpha ^* \beta \< f^*g \>
-\beta ^*\alpha \< g^*f \>+\beta\beta^* \< g^*g \>\\
&=|\alpha |^2\< f^*f \>-2\re(\alpha ^* \beta \< f^*g \>)
+|\beta|^2\< g^*g \>
\end{array}
\]
We now choose $\beta=\< f^*g \>$, and obtain for arbitrary
real $\alpha $ the inequality
\lbeq{f.8}
0\le \alpha ^2\< f^*f \>
-2\alpha |\< f^*g \>|^2+|\< f^*g \>|^2\< g^*g \>.
\eeq
The further choice $\alpha=\< g^*g \>$ gives
\[
0\le \< g^*g \>^2\< f^*f \>-\< g^*g \>|\< f^*g \>|^2.
\]
If $\< g^*g \>>0$, we find after division by $\< g^*g \>$ that (i) 
holds. And if $\< g^*g \>\le 0$ then $\< g^*g \>=0$ and we have 
$\< f^*g \>=0$ since otherwise a tiny $\alpha $ produces a negative
right hand side in \gzit{f.8}. Thus (i) also holds in this case.

(ii) Since $(f-\bar f)^*(g-\bar g)-(g-\bar g)^*(f-\bar f)=f^*g-g^*f$,
it is sufficient to prove the uncertainty relation for the case of
quantities $f,g$ whose expectation vanishes. In this case, (i) implies
\[
(\re \<f^*g\>)^2 +(\im \<f^*g\>)^2 =|\<f^*g\>|^2 \leq 
\< f^*f \>\< g^*g \> = \sigma(f)^2\sigma(g)^2.
\]
The assertion follows since $\re \<f^*g\>=\cov(f,g)$ and
\[
i\im \<f^*g\>=\shalf(\<f^*g\>-\<f^*g\>^*)=\shalf\<f^*g-g^*f\>.
\]

(iii) Again, it is sufficient to consider the case of
quantities $f,g$ whose expectation vanishes. Then
\lbeq{esig1}
\begin{array}{lll}
\sigma(f+g)^2 &=& \<(f+g)^*(f+g)\>
=\<f^*f\>+\<f^*g+g^*f\>+\<g^*g\>\\
&=& \sigma(f)^2+2\cov(f,g)+\sigma(g)^2,
\end{array}
\eeq
and \gzit{ecov1} follows. \gzit{ecov} is an immediate consequence of
(ii), and \gzit{esig} follows easily from \gzit{esig1} and 
\gzit{ecov}. Finally, \gzit{e.prodbound} is a consequence of 
\gzit{ecov} and Proposition \ref{p5.2}(iii).
\epf

In the classical case of commuting Hermitian quantities, the 
uncertainty relation just reduces to the well-known inequality 
\gzit{ecov} of classical statistics. For noncommuting Hermitian 
quantities, the uncertainty relation is stronger. In particular, we may
deduce from the commutation relation \gzit{ccr} for position $q$ and 
momentum $p$ {\sc Heisenberg}'s \cite{Hei,Rob} uncertainty relation
\lbeq{e6.unc0}
\sigma(q)\sigma(p)\geq \shalf\hbar.
\eeq
Thus {\em no ensemble exists 
where both $p$ and $q$ have arbitrarily small standard deviation}. 
(More general noncommuting Hermitian quantities $f,g$ may have 
{\em some} ensembles with $\sigma(f)=\sigma(g)=0$, namely among those 
with $\<fg\>=\<gf\>$.)

Putting $k=\bar p$ and $x=\bar q$ and taking expectations in 
\gzit{e6.unc} and using Proposition \ref{p5.2}(iii), we find another 
version of the uncertainty relation, implying again that $\sigma(p)$ 
and $\sigma(q)$ cannot be made simultaneously very small:
\lbeq{e6.unc1}
\Big(\frac{\sigma(p)}{\Delta p}\Big)^2
+\Big(\frac{\sigma(q)}{\Delta q}\Big)^2
\geq\frac{\hbar}{\Delta p \Delta q}.
\eeq
Heisenberg's relation \gzit{e6.unc0} follows from it by putting
$\Delta p = \sigma(p)$ and $\Delta q = \sigma(q)$.

We now derive a characterization of the quantities $f$ with vanishing
uncertainty, $\sigma(f)=0$; in classical probability theory these 
correspond to quantities (random variables) that have fixed values 
in every realization.

\begin{dfn}
We say a quantity $f$ {\bf vanishes} in the ensemble $\<\cdot\>$ if
\[
\<f^*f\>=0.
\]
\end{dfn}

\begin{thm}~

(i) $\sigma(f)=0$ iff $f-\<f\>$ vanishes.

(ii) If $f$ vanishes in the ensemble $\<\cdot\>$ then $\<f\>=0$.

(iii) The set $V$ of vanishing quantities satisfies
\[
f+g\in V~~~\mbox{if } f,g\in V,
\]
\[
fg\in V~~~\mbox{if $g\in V$ and $f\in \Ez$ is bounded},
\]
\[
f^2\in V~~~\mbox{if $f\in V$ is Hermitian}.
\]
\end{thm}

\bepf
(i) holds since $g=f-\<f\>$ satisfies $\<g^*g\>=\sigma(f)^2$.

(ii) follows from Proposition \ref{p5.2}(ii).

(iii) If $f,g\in V$ then $\<f^*g\>=0$ and $\<g^*f\>=0$ by the
Cauchy-Schwarz inequality, hence 
$\<(f+g)^*(f+g)\>=\<f^*f\>+\<g^*g\>=0$, so that $f+h\in V$.

If $g\in V$ and $f$ is bounded then 
\[
(fg)^*(fg)=g^*f^*fg\leq g^*\|f\|^2g=\|f\|^2g^*g
\]
implies $\<(fg)^*(fg)\>\leq\|f\|^2\<g^*g\>=0$, so that $fg\in V$.

And if $f\in V$ is Hermitian then $\<f^2\>=\<f^*f\>=0$, and, again
by Cauchy-Schwarz, $\<f^4\>\leq\<f^6\>\<f^2\>=0$, so that $f^2\in V$.
\epf

\bigskip
\section{Nonlocality} 
\label{nonlocality}

\hfill\parbox[t]{8.8cm}{\footnotesize

{\em As the heavens are higher than the earth, so are my ways higher 
than your ways and my thoughts than your thoughts.}

The {\sc LORD}, according to Isaiah, ca. 540 B.C. \cite{Isa} 

\bigskip
{\em Before they call I will answer; while they are still speaking 
I will hear.}

The {\sc LORD}, according to Isaiah, ca. 540 B.C. \cite{Isa2} 
}\nopagebreak

\bigskip
A famous feature of quantum physics is its intrinsic nonlocality,
expressed by so-called {\bf Bell inequalities} 
(cf. {\sc Bell} \cite{Bel}, {\sc Clauser \& Shimony} \cite{ClaS}). 
The formulation given here depends on the most orthodox part of 
quantum mechanics only; it does not, as is usually done, refer to 
hidden variables, and involves no counterfactual reasoning.

\begin{thm}\label{t.bell}
Let $f_k$ ($k=1,2,3,4$) be Hermitian quantities satisfying 
\lbeq{e.bell1}
f_k^2\leq 1~~~\mbox{for } k=1,2,3,4.
\eeq

(i) {\rm (cf. {\sc Cirel'son} \cite{Cir})} 
For every ensemble,
\lbeq{e.bellq}
|\<f_1f_2\>+\<f_3f_2\>+\<f_3f_4\>-\<f_1f_4\>|\leq 2\sqrt{2}.
\eeq

(ii) {\rm (cf. {\sc Clauser} et al. \cite{ClaHS})}
If, for odd $j-k$, the quantities $f_j$ and $f_k$ commute and are 
uncorrelated then
\lbeq{e.bellc}
|\<f_1f_2\>+\<f_3f_2\>+\<f_3f_4\>-\<f_1f_4\>|\leq 2.
\eeq

\end{thm}

\bepf
(i) Write $\gamma$ for the left hand side of \gzit{e.bellq}. Using
the Cauchy-Schwarz inequality and the easily verified inequality 
\[
\sqrt{\alpha}+\sqrt{\beta}\leq\sqrt{2(\alpha+\beta)}~~~
\mbox{for all } \alpha, \beta \geq 0, 
\]
we find
\[
\begin{array}{lll}
\gamma&=&|\<f_1(f_2-f_4)\>+\<f_3(f_2+f_4)\>| \\
&\leq&\sqrt{\<f_1^2\>\<(f_2-f_4)^2\>}+\sqrt{\<f_3^2\>\<(f_2+f_4)^2\>} \\
&\leq&\sqrt{\<(f_2-f_4)^2\>}+\sqrt{\<(f_2+f_4)^2\>} \\
&\leq&\sqrt{2(\<(f_2-f_4)^2\>+\<(f_2+f_4)^2\>)} 
       =\sqrt{4\<f_2^2+f_4^2\>}=\sqrt{8}. \\
\end{array}
\]

(ii) By Proposition \ref{p5.2}(ii), $v_k:=\<f_k\>$ satisfies
$|v_k|\leq 1$. If $f_j$ and $f_k$ commute and are uncorrelated for 
odd $j-k$ then Proposition \ref{p5.2}(iv) implies $\<f_jf_k\>=v_jv_k$ 
for odd $j-k$. Hence
\[
\begin{array}{lll}
\gamma&=&|v_1v_2+v_3v_2+v_3v_4-v_1v_4| =|v_1(v_2-v_4)+v_3(v_2+v_4)| \\
&\leq&|v_1|~|v_2-v_4|+|v_3|~|v_2+v_4| \leq |v_2-v_4|+|v_2+v_4| \\
&=&2\max(|v_2|+|v_4|)\leq 2. \\
\end{array}
\]
\epf

\begin{expl}\label{spinpair}
In $\Cz^{4 \times 4}$, the four monomial matrices $f_j$ defined by
\[
f_1x=\left(\begin{array}{r}x_3\\x_4\\x_1\\x_2\end{array}\right),~
f_2x=\left(\begin{array}{r}x_2\\x_1\\x_4\\x_3\end{array}\right),~
f_3x=\left(\begin{array}{r}x_1\\x_2\\-x_3\\-x_4\end{array}\right),~
f_4x=\left(\begin{array}{r}x_1\\-x_2\\x_3\\-x_4\end{array}\right)
\]
satisfy \gzit{e.bell1}, and $f_j$ and $f_k$ commute and are 
uncorrelated for odd $j-k$. It is easily checked that
in the pure ensemble defined by the vector 
\[
\psi=
\left(\begin{array}{r}
\alpha_1\\-\alpha_2\\ \alpha_2\\ \alpha_1
\end{array}\right),
~~~\alpha_{1,2}=\sqrt{\frac{2\pm\sqrt{2}}{8}},
\]
$\<f_1f_2\>=\<f_3f_2\>=\<f_3f_4\>=-\<f_1f_4\>=\half\sqrt{2}$. Hence
\gzit{e.bellq} holds with equality and \gzit{e.bellc} is violated.
Indeed, since $\<f_k\>=0$ for all $k$, we see that $f_j$ and $f_k$ are 
correlated for odd $j-k$.

On identifying 
\[
\left(\begin{array}{l}x_1\\x_2\\x_3\\x_4\end{array}\right)
=\left(\begin{array}{l}x_1~~x_2\\x_3~~x_4\end{array}\right)
\]
and defining the tensor product action $u\otimes v: x \mapsto uxv^T$, 
the matrices $f_j$ can be written in terms of the Pauli spin matrices 
\gzit{e.pauli} as
\[
f_1 = \sigma_1 \otimes 1,~~f_2 = 1 \otimes \sigma_1,~~
f_3 = \sigma_3 \otimes 1,~~f_4 = 1 \otimes \sigma_3.
\]

If we interpret the two terms in a tensor product as quantities related to two spatially separated fermion particles $A$ and $B$, we conclude 
that there are pure ensembles in which the components of the spin 
vectors of two fermion particles are correlated, no matter how far 
apart the two particles are placed. 

\end{expl}

Such {\em nonlocal correlations} of certain quantum ensembles are
an enigma of the microscopic world that, being experimentally 
confirmed, cannot be removed by any interpretation of quantum
mechanics.
(See {\sc Bell} \cite{Bel} for the original Bell inequality,
{\sc Pitowsky} \cite{Pit} for a treatise on Bell inequalities,
and {\sc Aspect} \cite{Asp}, {\sc Clauser \& Shimony} \cite{ClaS}, 
{\sc Tittel} et al. \cite{TitBG} for experiments
verifying the violation of \gzit{e.bellc}.)

\section{Probability} 
\label{probability}

\hfill\parbox[t]{8.8cm}{\footnotesize

{\em Enough, if we adduce probabilities as likely as 
any others; for we must remember that I who am the speaker, and you 
who are the judges, are only mortal men, and we ought to accept the 
tale which is probable and enquire no further.}

Plato, ca. 367 B.C. \cite{Pla} 
}\nopagebreak

\bigskip
The interpretation of probability has been surrounded by philosophical
puzz\-les for a long time. {\sc Fine} \cite{Fin} is probably still the 
best discussion of the problems involved; {\sc Hacking} \cite{Hac}
gives a good account of its early history. (See also 
{\sc Home and Whitaker} \cite{HomW}.) Our definition generalizes
the classical intuition of probabilities as weights in a weighted 
average and is modeled after the formula for finite 
probability theory in Example \ref{ex.5.3}(i).

In the special case when a well-defined counting process may be 
associated with the statement whose probability is assessed, our 
exposition supports the conclusion of {\sc Drieschner} \cite[p.73]{Dri},
{\em ``probability is predicted relative frequency''}
(German original: ``Wahrscheinlichkeit ist vorausgesagte relative 
H\"au\-fig\-keit''). More specifically, we assert that, 
{\em for counting events, the probability carries the information of 
expected relative frequency} (see Theorem \ref{t1.5}(iii) below). 

To make this precise we need a precise concept of independent events 
that may be counted. To motivate our definition, assume that we look at
times $t_1,\dots,t_N$ for the presence of an event of the sort we want
to count. We introduce quantities $e_l$ whose value is the amount
added to the counter at time $t_l$. For correct counting, we need
$e_l\approx 1$ if an event happened at time $t_l$, and $e_l\approx 0$
otherwise; thus $e_l$ should have the two possible values $0$ and $1$ 
only. Since these numbers are precisely the Hermitian idempotents
among the constant quantities, this suggests to identify events with 
general Hermitian idempotent quantities. 

In addition, it will be useful to have the more general concept of 
`effects' for more fuzzy, event-like things. 

\begin{dfn}~\nopagebreak

(i) A quantity $e \in \Ez$ satisfying $0\leq e\leq 1$ is called an 
{\bf effect}. The number 
$\<e\>$ is called the {\bf probability} of the effect $e$.
Two effects $e,e'$ are called {\bf independent} in an ensemble 
$\< \cdot \>$ if they commute and satisfy
\[
  \< ee' \> = \< e \> \< e' \>.
\]

(ii) A quantity $e \in \Ez$ satisfying $e^2 = e = e^*$ is called an 
{\bf event}. Two events $e,e'$ are called {\bf disjoint} if 
$ee'=e'e=0$.

(iii) An {\bf alternative} is a family $e_l$ ($l\in L$) of effects such
that 
\[
\sum_{l\in L} e_l \leq 1.
\]
\end{dfn}

\begin{prop}~

(i) Every event is an effect.

(ii) The probability of an effect $e$ satisfies $0\leq\<e\>\leq 1$.

(iii) The set of all effects is convex and closed in the uniform 
topology.

(iv) Any two events in an alternative are disjoint.

\end{prop}

\bepf
(i) holds since $0\leq e^*e=e^2=e$ and 
$0\leq(1-e)^*(1-e)=1-2e+e^2=1-e$. 

(ii) and (iii) follow easily from Proposition \ref{p5.2}.

(iv) If $e_k,e_l$ are events in an alternative then $e_k\leq 1-e_l$
and
\[
(e_ke_l)^*(e_ke_l)=e_l^*e_k^*e_ke_l=e_l^*e_k^2e_l=e_l^*e_ke_l
\leq e_l^*(1-e_l)e_l=0.
\]
Hence $e_ke_l=0$ and $e_le_k=e_l^*e_k^*=(e_ke_l)^*=0$.
\epf

Note that we have a well-defined notion of probability though the
concept of a probability distribution is absent. It is neither needed 
nor definable in general. Nevertheless, the theory contains classical
probability theory as a special case.

\begin{expls}\label{ex1.4}~

(i) {\bf Classical probability theory.} 
In classical probability theory, quantities are usually called 
{\bf random variables}; they belong to the Q-algebra $B(\Omega)$ of 
measurable complex-valued functions on a measurable set $\Omega$. 

The characteristic function $e = \CHI_M$ of any measurable subset $M$ 
of $\Omega$ (with $\CHI_M (\omega ) =1$ if $\omega  \in M$, 
$\CHI_M (\omega )=0$ otherwise) is an event.
A family of characteristic functions $\CHI_{M_l}$ form an alternative 
iff their supports $M_l$ are pairwise disjoint. 

Effects are the measurable functions $e$ with values in $[0,1]$; they 
can be considered as `characteristic functions' of a fuzzy set where 
$\omega \in\Omega$ has $e(\omega )$ as degree of membership (see, e.g., 
{\sc Zimmermann} \cite{Zim}).

For many applications, the algebra $B(\Omega)$ is too big, and 
suitable subalgebras $\Ez$ are selected on which the relevant 
ensembles can be defined as integrals with respect to suitable positive
measures.

(ii)  {\bf Quantum probability theory.} 
In the algebra of bounded linear operators on a Hilbert space 
$\Hz$, every unit vector $\phi \in \Hz$ gives rise to an 
{\bf elementary event} $e_\phi = \phi\phi ^*$.
A family of elementary events $e_{\phi_l}$ form an alternative 
iff the $\phi_l$ are pairwise orthogonal. The probability of an
elementary event $e_\phi$ in an ensemble corresponding to the
unit vector $\psi$ is 
\lbeq{e.sqprob}
\<e_\phi\>=\psi^*e_\phi\psi=\psi^*\phi\phi^*\psi=|\phi^*\psi|^2.
\eeq
This is the well-known {\bf squared probability amplitude} formula,
traditionally interpreted as the probability that after preparing a 
pure ensemble in `state' $\psi$, an ideal measurement causes a 
`state reduction' to the new pure `state' $\phi$. Note that our
interpretation of $|\phi^*\psi|^2$ is completely within the formal
framework of the theory and completely independent of the 
measurement process.

Further, nonelementary quantum events are orthogonal projectors to 
subspaces. The effects are the Hermitian operators $e$ with spectrum 
in $[0,1]$.
\end{expls}

\begin{thm}\label{t1.5}~

(i) For any effect $e$, its {\bf negation} $\neg e = 1 - e$ is an
effect with probability 
\[
  \< \neg e \> = 1 - \< e \>;
\]
it is an event if $e$ is an event.

(ii) For commuting effects $e, e'$, the quantities
\[
  e \wedge e' = ee' ~~~(e \mbox{ \bf and } e'),
\]
\[
  e \vee e' = e + e' - ee' ~~~(e \mbox{ \bf or } e')
\]
are effects whose probabilities satisfy
\[
  \< e \wedge e' \> + \< e \vee e' \> = \< e \> + \< e' \>;
\]
they are events if $e, e'$ are events. Moreover,
\[
\<e\wedge e'\> = \<e\>\<e'\>~~~\mbox{for independent effects } e, e'.
\]

(iii) For a family of effects $e_l$ $(l=1, \ldots , N)$ with
constant probability $\< e_l \> = p$, the {\bf relative frequency}
\[
  q := \frac{1}{N} \D \sum ^N _{l=1} e_l
\]
satisfies
\[
  \< q \> =p.
\]
(iv) For a family of independent events of probability $p$,
the uncertainty 
\[
\sigma (q) = \sqrt{ \frac{ p(1-p)}{N}}
\]
of the relative frequency becomes arbitrarily small as $N$ becomes 
sufficiently large {\bf (weak law of large numbers)}.
\end{thm}

\bepf
(i) $\neg e$ is an effect since $0\leq 1-e\leq1$, and its probability
is $\< \neg e \> = \< 1-e \> = 1 - \< e \>$.
If $e$ is an event then clearly $\neg e$ is Hermitian, and 
$(\neg e)^2=(1-e)^2=1-2e+e^2=1-e=\neg e$. Hence $\neg e$ is an event.

(ii) Since $e$ and $e'$ commute, $e \wedge e'=ee'=e^2e'=ee'e$.
Since $ee'e\geq0$ and $ee'e\leq ee=e\leq1$, we see that $e \wedge e'$
is an effect. Therefore, 
$e \vee e'=e+e'-ee'=1-(1-e)(1-e')=\neg (\neg e\wedge \neg e')$ 
is also an effect. The assertions about expectations are immediate.
If $e,e'$ are events then $(ee')^* =e'^* e^* =e'e=ee'$, hence 
$ee'$ is Hermitian; and it is idempotent since 
$(ee')^2=ee'ee'=e^2e'^2=ee'$. Therefore $e \wedge e'=ee'$ is an
event, and $e \vee e'=\neg (\neg e\wedge \neg e')$ is an event, too.

(iii) This is immediate by taking the expectation of $q$.

(iv) This follows from Theorem \ref{t.weaklaw} since
$\<e_k^2\>=\<e_k\>=p$ and 
\[
\sigma(e_k)^2= \< (e_k-p)^2\> =\< e_k^2\> -2p\< e_k\> +p^2 
=p-2p^2+p^2=p(1-p).
\]
\epf

We remark in passing that, with the operations $\wedge,\vee,\neg$, 
the set of events in any {\em commutative} subalgebra of $\Ez$ 
forms a Boolean algebra; see {\sc Stone} \cite{Sto}. Traditional
quantum logic (see, e.g., {\sc Birkhoff \& von Neumann} \cite{BirN}, 
{\sc Pi\-tow\-sky} \cite{Pit}, {\sc Svozil} \cite{Svo}) discusses the 
extent to which this can be generalized to the noncommutative case. 
We shall make no use of quantum logic; the only logic used is 
classical logic, applied to well-defined assertions about quantities.
However, certain facets of quantum logic related to so-called 
`hidden variables' are discussed from a different point of view in the 
next section.

The set of effects in a commutative subalgebra is {\em not} a 
Boolean algebra. Indeed, $e \wedge e \neq e$ for effects $e$ that are 
not events. In fuzzy set terms, if $e$ codes the answer to the question 
`(to which degree) is statement $S$ true?' then $e \wedge e$ codes the 
answer to the question `(to which degree) is statement $S$ really 
true?', indicating the application of more stringent criteria for 
truth.

For noncommuting effects, `and' and `or' ar undefined. One might 
think of $\shalf(ee'+e'e)$ as a natural definition for $e \wedge e'$;
however, this expression need not be an event, as the simple example
\[
e=\left(\bary{cc}1&0\\0&0 \end{array}\right),~~
e'=\half\left(\bary{cc}1&1\\1&1 \end{array}\right),~~~~
\half(ee'+e'e)=\frac{1}{4}\left(\bary{cc}2&1\\1&0 \end{array}\right)
\]
shows.

\newpage
\section{States} \label{states}

\hfill\parbox[t]{8.8cm}{\footnotesize

{\em For example, nobody doubts that at any given time the center of 
mass of the Moon has a definite position, even in the absence of any 
real or potential observer.}

Albert Einstein \cite{Ein2} 
}\nopagebreak

\bigskip
States formalize the objective properties that physical systems 
possess. We consider properties (the `beables' of 
{\sc Bell} \cite{Bel2}) to be assignments of complex numbers $v(f)$ to 
certain quantities $f$. 

The specification of which states correspond to physical
systems is part of the interpretation problem of quantum mechanics.
Different schools use different proposals but, due to the lack of 
experimental tests, no agreement has been reached.
We therefore demand only minimal requirements shared by all reasonable
concepts of states, and independent on any a priori relations to 
(as yet undefined) measurement.

We discuss the constraints imposed on sharpness, 
a desirable property of Hermitian quantities.
In this way we find an answer to the question: Assuming there
is an objective reality behind quantum physics, what form can it take?

Since not all states assign properties to all quantities, we need a 
symbol `?' that indicates an unspecified (and perhaps undefined) value. 
Operations involving ? give ? as a result, with exception of the
rule
\[
0?=?0=0.
\]

\begin{dfn}~

(i) A {\bf state} is a mapping $v : \Ez \rightarrow \Cz \cup \{ ? \}$
such that

(S1) $v(\alpha+\beta f)=\alpha+\beta  v(f)$~~~ if $\alpha, \beta\in\Cz$,

(S2) $v (f)\in\Rz\cup\{?\}$ ~~~if $f$ is Hermitian.

$v(f)$ is called the {\bf reference value} of $f$ in state $v$.
$\Ez_v:= \{ f\in \Ez \mid v (f)\in \Cz \}$
denotes the set of quantities with definite values in state $v$.

(ii) A set $E$ of Hermitian quantities is called {\bf sharp} in state 
$v$ if, for $f,g \in E$ and $\lambda\in\Rz$,,

(SQ0) $\Rz \subseteq E,~~~v(f)\in\Rz$,

(SQ1) $f^2 \in E,~~~ v(f^2) = v(f)^2$,
 
(SQ2) $f^{-1}\in E$,~~~$v(f^{-1})=v(f)^{-1}$ ~~~if $f$ is invertible,

(SQ3) $f\pm g \in E,~ v(f+\lambda g) = v(f) + \lambda v(g)$ 
~~~if $f,g$ commute.

A quantity $f$ is called {\bf sharp} in $v$ if $\re f$ and $\im f$ 
commute and belong to some set that is sharp in state $v$.

\end{dfn}

Thus, sharp quantities behave with respect to their reference values
precisely as numbers would do. In particular, sharp quantities are 
normal by Corollary \ref{c1.3}.

While having a well-defined reference value guarantees objectivity and 
hence observer-independent reality, sharpness is a matter not of 
objectivity but one of point-like behavior. 

\begin{expls}\label{ex.classquant}~

(i) {\bf Classical mechanics.} 
Classical $N$-particle mechanics is described by a {\em phase space} 
$\Omega_{cl}$, the direct product of $\Rz^N\times\Rz^N$ and a compact 
manifold describing internal particle degrees of freedom. 
$\Ez$ is a subalgebra of the algebra $B (\Omega_{cl})$ of
Borel measurable functions on {\bf phase space} $\Omega_{cl}$.

A {\bf classical point state} is defined for each 
$\omega\in\Omega_{cl}$ by 
\[
  v_\omega (f) := 
   \left\{ \bary{ll}
     f(\omega) & \mbox{if  $f$ is continuous at $\omega$},\\
     ? & \mbox{otherwise.}
   \eary \right.
\]
In a classical point state $v$, all $f\in\Ez_v$ are sharp (and normal).

(ii) {\bf Nonrelativistic quantum mechanics.}
Nonrelativistic quantum mechanics of $N$ particles is described by a 
Hilbert space $\Hz=L^2(\Omega_{qu})$, where $\Omega_{qu}$ is the 
direct product of $\Rz^N$ and a finite set that takes care of spin, 
color, and similar indices. $\Ez = \Ez_2 (\Omega)$ is the algebra of 
bounded linear operators on $\Hz$. 
(If unbounded operators are considered, $\Ez$ is instead an algebra of 
linear operators in the corresponding Schwartz space, but for this 
example, we don't want to go into technical details.) 

The Copenhagen interpretation is the most prominent, and at the same 
time the most restrictive interpretation of quantum mechanics. 
It assigns definite values only to quantities in an eigenstate.
A {\bf Copenhagen state} is defined for each 
$\psi \in \Hz\setminus \{0\}$ by
\[
  v _\psi (f) := 
   \left\{ \bary{ll}
     \lambda & \mbox{if } f \psi = \lambda \psi, \\
     ? & \mbox{otherwise.}
   \eary \right.
\]
In a Copenhagen state $v$, all normal $f\in\Ez_v$ are sharp.

\end{expls}~

Our first observation is that numbers are their own reference values, 
and that sharp events are dichotomic -- their only possible reference 
values are $0$ and $1$.

\begin{prop}~

(i)~ $v(\alpha)=\alpha$~ if $\alpha \in \Cz$.

(ii) If $e$ is a sharp event then $v(e)\in\{0,1\}$.
\end{prop}

\bepf
(i) is the case $\beta=0$ of (S1), and (ii) holds since
in this case, (SQ1) implies $v(e)=v(e^2)=v(e)^2$.
\epf

\begin{prop}
If the set $E$ is sharp in the state $v$ then
\lbeq{e.s1}
fg\in E,~~v(fg) = v(f)v(g)~~~\mbox{if $f,g \in E$ commute},
\eeq
\lbeq{e.s2}
\alpha+\beta f \in E,~~v(\alpha+\beta f)=\alpha+\beta v(f)
~~~\mbox{if } f \in E,\alpha,\beta\in\Rz.
\eeq
\end{prop}

\bepf
If $f,g \in E$ commute then $f\pm g \in E$ by (SQ3). By (SQ1),
$(f\pm g)^2\in E$ and $v((f\pm g)^2)=v(f\pm g)^2$. By (SQ3),
$fg=((f+g)^2-(f-g)^2)/4$ belongs to $E$ and satisfies
\[
\begin{array}{lll}
4v(fg)&=&v((f+g)^2)-v((f-g)^2)=v(f+g)^2-v(f-g)^2\\
&=&(v(f)+v(g))^2-(v(f)-v(g))^2=4v(f)v(g).
\end{array}
\]
Thus \gzit{e.s1} holds, and \gzit{e.s2} 
follows from \gzit{e.s1}, (SQ0) and (SQ3). 
\epf

\bigskip
One of the nontrivial traditional {\em postulates} of quantum mechanics,
that the possible values a sharp quantity $f$ may take are the 
elements of the spectrum $\spec f$ of $f$, is a {\em consequence} of 
our axioms. 

\begin{thm}\label{t.spec}
If a Hermitian quantity $f$ is sharp with respect to $v$, and 
$v(f)=\lambda$ then: 

(i) $\lambda-f$ is not invertible. 

(ii) If there is a polynomial $\pi(x)$ such that $\pi(f)=0$
then $\lambda$ satisfies $\pi(\lambda)=0$. In particular,
if $f$ is a sharp event then $v(f)\in \{0,1\}$.

(iii) If $\Ez$ is finite-dimensional then there is a quantity $g\neq 0$
such that $fg=\lambda g$, i.e., $\lambda$ is an eigenvalue of $f$.

\end{thm}

\bepf
Note that $\lambda$ is real by (SQ0).

(i) If $g:=(\lambda-f)^{-1}$ exists then by \gzit{e.s2} and (SQ2),
$\lambda-f,g\in E$ and
\[
v(\lambda-f)v(g)=v((\lambda-f)g)=v(1)=1,
\]
contradicting $v(\lambda-f)=\lambda-v(f)=0$.

(ii) By polynomial division we can find a polynomial $\pi_1(x)$ such 
that $\pi(x)=\pi(\lambda)+(x-\lambda)\pi_1(x)$. If $\pi(\lambda)\neq 0$,
$g:=-\pi_1(f)/\pi(\lambda)$ satisfies 
\[
(\lambda-f)g=(f-\lambda)\pi_1(f)/\pi(\lambda)
=(\pi(\lambda)-\pi(f))/\pi(\lambda)=1,
\]
hence $\lambda-f$ is invertible with inverse $g$, contradiction. 
Hence $\pi(\lambda)=0$. In particular, this applies to an event with 
$\pi(x)=x^2-x$; hence its possible reference values are zeros of 
$\pi(x)$, i.e., either $0$ or $1$.

(iii) The powers $f^k$ ($k=0,\dots, \dim \Ez$) must be linearly 
dependent; hence there is a polynomial $\pi(x)$ such that $\pi(f)=0$.
If this is chosen of minimal degree then $g:=\pi_1(f)$ is nonzero
since its degree is too small. Since
$0=\pi(\lambda)=\pi(f)+(f-\lambda)\pi_1(f)=(f-\lambda)g$, we have
$fg=\lambda g$.
\epf

When $\Ez$ is a $C^*$-algebra, the spectrum of $f\in \Ez$ is defined 
as the set of complex numbers $\lambda$ such that $\lambda-f$ has no 
inverse (see, e.g., \cite{Ric}). Thus in this case, part (i) of the 
theorem implies that all numerical values a sharp quantity $f$ can 
take belong to the spectrum of $f$. This covers both the case of 
classical mechanics and that of nonrelativistic quantum mechanics.

However, in general, one cannot hope that {\em every} Hermitian 
quantity is sharp. Indeed, it was shown already by 
{\sc Kochen \& Specker} \cite{KocS} that there is a finite set of 
events in $\Cz^{3 \times 3}$ (and hence in $\Cz^{n \times n}$ for 
all $n\geq 3$) for which any assignment of reference values leads to a 
contradiction with the sharpness conditions. We give a slightly less 
general result that is much easier to prove.

\begin{thm}\label{t.nocons}
{\rm (cf. {\sc Mermin} \cite{Mer}, {\sc Peres} \cite{Per2})}

There is no state with a sharp set of quantities containing four
Hermitian quantities $f_j$ ($j=1,2,3,4$) satisfying $f_j^2=1$ and
\lbeq{e.nocons}
f_jf_k=\left\{ \begin{array}{rl}
-f_kf_j &\mbox{if } j-k=\pm 2,\\
f_kf_j &\mbox{otherwise}.
\end{array}\right.
\eeq
\end{thm}

\bepf
Let $E$ be a set containing the $f_j$. If $E$ is sharp in the state $v$
then $v_j=v(f_j)$ is a number, and $v_j^2=v(f_j^2)=v(1)=1$ implies 
$v_j \in\{-1,1\}$. In particular,
$v_0:=v_1 v_2 v_3 v_4\in\{-1,1\}$.
By \gzit{e.s1}, $v(f_j f_k)=v_j v_k$ if $j,k\neq \pm 2$. 
Since $f_1f_2$ and $f_3f_4$ commute,
$v(f_1f_2f_3f_4)=v(f_1f_2) v(f_3f_4)=v_1v_2v_3 v_4
=v_0$, and since $f_1f_4$ and $f_2f_3$ commute,
$v(f_1f_4f_2f_3)=v(f_1f_4)v(f_2f_3) v_1v_4v_2 v_3
=v_0$. Since $f_1f_4f_2f_3=-f_1f_2f_3f_4$, this gives 
$v_0=-v_0$, hence the contradiction $v_0=0$.
\epf

\begin{expl}\label{ex.peres}
The $4 \times 4$-matrices $f_j$ defined in Example \ref{spinpair} 
satisfy the required relations. In particular, there cannot be a state 
in which all components of the spin vectors of two fermions are sharp.

This is the basic reason underlying a number of well-known arguments 
against so-called local hidden variable theories, which assume that 
{\em all} Hermitian quantities are sharp. 
(See {\sc Bernstein} \cite{Ber}, {\sc Eberhard} \cite{Ebe}, 
{\sc Greenberger} et al. \cite{GreHS,GreHZ}, 
{\sc Hardy} \cite{Har,Har2}, {\sc Mermin} \cite{Mer,Mer2}, 
{\sc Peres} \cite{Per,Per2}, {\sc Vaidman} \cite{Vai}). 
For a treatment in terms of quantum logic, see {\sc Svozil} \cite{Svo}.

\end{expl}

Sharp quantities always satisfy a {\bf Bell inequality} analogous to 
inequality \gzit{e.bellc} for uncorrelated quantities:

\begin{thm}\label{t.bells}
Let $v$ be a state with a sharp set of quantities containing four
Hermitian quantities $f_j$ ($j=1,2,3,4$) satisfying $f_j^2=1$
and $[f_j,f_k]=0$ for odd $j-k$. Then 
\lbeq{e.bellcs}
|v(f_1f_2)+v(f_2f_3)+v(f_3f_4)-v(f_1f_4)|\leq 2.
\eeq

\end{thm}

\bepf
Let $v_k:=v(f_k)$. Then (SQ2) implies $v_k^2=v(f_k^2)=v(1)=1$, and
since equation \gzit{e.s1} implies  $v(f_jf_k)=v_jv_k$ for odd
$j-k$, we find
\[
\begin{array}{lll}
\gamma&=&|v_1v_2+v_2v_3+v_3v_4-v_1v_4| \\
&=&|v_1(v_2-v_4)+v_3(v_2+v_4)| \\
&\leq&|v_1|~|v_2-v_4|+|v_3|~|v_2+v_4| \\
&\leq&|v_2-v_4|+|v_2+v_4| \\
&=&2\max(|v_2|+|v_4|)\leq 2. \\
\end{array}
\]
\epf

Note, however, that Example \ref{ex.peres} already implies that the 
sharpness assumption in this theorem (and in other derivations of 
Bell inequalities for local hidden variable theories; see, e.g., the 
treatise {\sc Pitowsky} \cite{Pit}) fails not only in special entangled 
ensembles such as that exhibited in Example \ref{spinpair} but must 
fail independent of any special preparation.

While the above results show that one cannot hope to find 
quantum states in which all Hermitian quantities are sharp, results of
{\sc Clifton \& Kent} \cite{CliK} imply that one can achieve sharpness 
in $\Ez=\Cz^{n\times n}$ at least for a dense subset of Hermitian 
quantities.

\section{States without squaring rule}

\hfill\parbox[t]{6.5cm}{\footnotesize

{\em But if we have food and clothing, we will be content with that.}

St. Paul, ca. 60 A.D. \cite{Pau3} 

}\nopagebreak

\bigskip
Since sharpness cannot be achieved for all Hermitian quantities, we 
discuss the relevance of the sharpness assumption.

The chief culprit among the sharpness assumptions seems to be the
squaring rule (SQ1) from which the product rule \gzit{e.s1} was
derived. Indeed, the squaring rule (and hence the product rule) already 
fails in a simpler, classical situation, namely when considering 
weak limits of highly oscillating functions, 
For example, consider the family of functions $f_k$ defined on $[0,1]$
by $f_k(x)=\alpha$ if $\lfloor kx \rfloor$ is even and $f_k(x)=\beta$ 
if $\lfloor kx \rfloor$ is odd. Trivial integration shows that  
the weak-$^*$ limits are $\lim f_k=\shalf(\alpha+\beta)$ and 
$\lim f_k^2=\shalf(\alpha^2+\beta^2)$, and these do not satisfy the 
expected relation $\lim f_k^2= (\lim f_k)^2$. Such weak limits of 
highly oscillating functions lead to the concept of a 
{\em Young measure}, which is of relevance in the calculus of 
variation of nonconvex functionals and in the physics of metal 
microstructure. See, e.g., {\sc Roubicek} \cite{Rou}.

More insight from the classical regime comes from realizing that 
reference values are a microscopic analogue of similar macroscopic 
constructions. 

For example, the center of mass, the mass-weighted average of the 
positions of the constituent particles, serves in classical mechanics 
as a convenient reference position of an extended object. It defines a 
point in space with a precise and objective physical meaning. 
The object is near this reference position, within an uncertainty 
given by the diameter of the object. Similarly, a macroscopic object 
has a well defined reference velocity, the mass-weighted average of 
the velocities of the constituent particles.

Thus, if we define an algebra $\Ez$ of `intensive' macroscopic 
mechanical quantities, given by all (mass-independent and sufficiently 
nice) functions of time $t$, position $q(t)$, velocity $\dot q(t)$ 
and acceleration $\ddot q(t)$, the natural reference value 
$v_{mac}(f)$ for a quantity $f$ is the mass-weighted average of the 
$f$-values of the constituent particles (labeled by superscripts $a$),
\[
v_{mac}(f)= 
\sum_a m^a f(t, q^a(t),\dot q^a(t), \ddot q^a(t))\Big/\sum_a m^a.
\]
This reference value behaves correctly under aggregation, if on the 
right hand side the reference values of the aggregates are substituted,
so that it is independent of the details of how the object is split 
into constituents. Moreover, $v=v_{mac}$ has nice properties: 
{\bf unrestricted additivity},

(SL) $v(f+g) = v(f) + v(g)$ ~~~if $f,g \in \Ez$,

and {\bf monotony},

(SM) $f\geq g \implies v(f)\geq v(g)$.

However, neither position nor velocity nor acceleration is a sharp 
quantity with respect to $v_{mac}$ since (SQ1) and (SQ2) fail. 
Note that deviations from the squaring rule make physical sense; 
for example, $v_{mac}(\dot q^2)-v_{mac}(\dot q)^2$ is 
(in thermodynamic equilibrium) proportional to the temperature of the 
system. 

From this perspective, and in view of Einstein's quote at the 
beginning of section \ref{states}, demanding the squaring rule for a 
reference value is unwarranted since it does not even hold in this 
classical situation.

\bigskip
Once the squaring rule (and hence sharpness) is renounced as a 
requirement for definite reference values, the arena is free for 
interpretations that use reference values defined for {\em all} 
quantities, and thus give a satisfying realistic picture of quantum 
mechanics. In place of the lost multiplicative properties we may now 
require unrestricted additivity (SL) without losing interesting 
examples. 

For example, the `local expectation values' of Bohmian mechanics 
({\sc Bohm} \cite{Boh}) have this property, if the prescription given 
for Hermitian quantities in {\sc Holland} \cite[(3.5.4)]{Hol} is 
extended to general quantities, using the formula
\[
v(f) :=v(\re f) +iv(\im f)
\]
which follows from (SL). Such {\bf Bohmian states} have, by design, 
sharp positions at all times. However, they lack 
desirable properties such as monotony (SM), and they display other 
counterintuitive behavior (see, e.g., {\sc Neumaier} \cite{Neu.bohm} 
and its references). 

A much more natural proposal is to {\em require that each state is an
ensemble}. Then (SL) and (SM) hold, and one even has a replacement
for the multiplicative properties. Indeed, for such 
{\bf ensemble states}, it follows from \gzit{e.prodbound} that 
there is an uncertainty measure 
\lbeq{e.uncmeas}
\Delta f =\sqrt{v(f^2)-v(f)^2}
\eeq
associated with each Hermitian quantity $f$ such that
\lbeq{e.prods}
|v(fg)-v(f)v(g)|\leq\Delta f\Delta g
~~~\mbox{for commuting Hermitian } f,g.
\eeq
Thus the product rule (and in particular the squaring rule) holds in an
approximate form. 

For quantities with small uncertainty $\Delta f$, we have essentially 
classical (nearly sharp) behavior. Im particular,
by the weak law of large numbers, Theorem \ref{t.weaklaw}, averages 
over many uncorrelated commuting quantities of the same kind have small
uncertainty and hence are nearly classical. In particular, this holds
for the quantities considered in statistical mechanics, and 
{\em explains the emergence of classical properties for macroscopic 
systems}.

Indeed, in statistical mechanics, classical values for observables 
are traditionally defined as expectations, and the concept of ensemble 
states with objective reference values for all quantities simply 
extends this downwards to the quantum domain. 

With the interpretation that the only states realized in quantum 
mechanics are ensembles states, 
{\em quantum objects are inherently extended objects}, and realizing 
this reduces the riddles the interpretation of the microworld poses 
when instead pointlike (sharp) properties are imagined. 

\begin{expls}~

(i) {\bf The ground state of hydrogen.}
The uncertainty $\Delta q$ of position (defined by 
interpreting \gzit{e.uncmeas} for the vector $q$ in place of the 
scalar $f$) in the ground state of hydrogen is 
$\Delta q=\sqrt{3} r_0$ (where $r_0=5.29\cdot 10^{-11}\fct{m}$ is the 
Bohr radius of a hydrogen atom), slightly larger than the reference 
radius $v(r)=v(|q-v(q)|)=1.5 r_0$.

(ii) {\bf The center of mass of the Moon.}
The Moon has a mass of $m_{\fns{Moon}}=7.35\cdot 10^{22} \fct{kg}$,
Assuming the Moon consists mainly of silicates, we may take the
average mass of an atom to be about 20 times the proton mass
$m_{\fns{p}}=1.67 \cdot 10^{-27} \fct{kg}$. Thus the Moon contains 
about $N=m_{\fns{Moon}}/20m_{\fns{p}}=2.20\cdot 10^{48}$ atoms.
In the rest frame of an observer standing on the Moon, the objective 
uncertainty of an atom position (due to the thermal motion of the 
atoms in the Moon) may be taken to be a small multiple of the Bohr 
radius $r_0$. Assuming that the deviations from the 
reference positions are uncorrelated, we may use \gzit{e.sigN}
to find as uncertainty of the position of the center of mass of the 
Moon a small multiple of $r_0/\sqrt{N}=3.567\cdot 10^{-35}\fct{m}$. 
Thus the center of mass of the Moon has a definite objective 
position, sharp within the measuring accuracy of many generations to 
come. 
\end{expls}

Ensemble states provide an elegant solution to the reality problem, 
confirming the insistence of the orthodox Copenhagen interpretation 
on that there is nothing but ensembles, while avoiding their elusive 
reality picture. It also conforms to {\sc Ockham}'s razor 
\cite{Ock, HofMC}, {\em frustra fit per plura quod potest fieri per 
pauciora}, that we should not use more degrees of freedom than are 
necessary to explain a phenomenon.

Quantum reality with reference values defined by ensemble states is as 
well-behaved and objective as classical macroscopic reality with 
reference values defined by a mass-weighted average over constituent 
values, and lacks sharpness (in the sense of our definition) to the 
same extent as classical macroscopic reality. 

Moreover, classical point states are ensemble states, and whenever a 
Copenhagen state assigns a numerical value to a quantity, the 
corresponding pure ensemble state assigns the same value to it. Thus
both classical mechanics and the orthodox interpretation of quantum 
mechanics are naturally embedded in the ensemble state interpretation.

The logical riddles of quantum mechanics (see, e.g., 
{\sc Svozil} \cite{Svo}) find their explanation in 
the fact that most events are unsharp in a given ensemble state, so 
that their objective reference values are no longer dichotomic but 
may take arbitrary values in $[0,1]$, by (SM).

The arithmetical riddles of quantum mechanics (see, e.g., 
{\sc Schr\"odinger} \cite{Sch}) find their explanation in the fact 
that most Hermitian quantities are unsharp in a given ensemble state, 
so that their objective reference values are no longer eigenvalues but 
may take arbitrary values in the convex hull of the eigenvalues.

The geometric riddles of quantum mechanics -- e.g., in the double 
slit experiment ({\sc Bohr} \cite{Bohr}, {\sc Wootters \& Zurek} 
\cite{WooZ}) and in EPR-experiments ({\sc Aspect} \cite{Asp}, 
{\sc Clauser \& Shimony} \cite{ClaS}) -- do not disappear. 
But they remain within the magnitudes predicted by reference radii
and uncertainties, hence require no special interpretation in the 
microscopic case. They simply demonstrate that particles are 
intrinsically extended and that electrons cannot be regarded
as pointlike. (For photons, this is known to be the case also for
different reasons, namely the nonexistence of a position operator 
with commuting components; see, e.g., {\sc Strnad} \cite{Str}, 
{\sc Mandel \& Wolf} \cite[Chapter 12.11]{ManW}, 
{\sc Newton \& Wigner} \cite{NewW}, {\sc Pryce} \cite{Pry}, but
cf. {\sc Hawton} \cite{Haw}.) 

Moreover, when considering quantum mechanical phenomena that violate 
our geometric intuition, one should bear in mind two similar 
violations of a naive geometric picture for the center of mass, 
Einstein's prototype example for a definite and objective property of 
macroscopic systems:

First, though it is objective, the center of mass is nevertheless 
a fictitious point, not visibly distinguished in reality; for nonconvex 
objects it may even lie outside the object! 
Second, the center of mass follows a well-defined, objective path, 
though this path need not conform to the visual path of the object; 
this can be seen by pushing a long, elastic cylinder through a strongly 
bent tube. 

\bigskip
All these considerations are independent of the measurement problem.
To investigate how measurements of classical macroscopic quantities
(i.e., expectations of quantities with small uncertainty related to a
measuring device) correlate with reference values of a microscopic
system interacting with the device requires a precise definition of a 
measuring device and of the behavior of the combined system under the
interaction (cf. the treatments in {\sc Busch} et 
al. \cite{BusGL,BusLM}, {\sc Giulini} et al. \cite{GiuJK},
{\sc Mittelstaedt} \cite{Mit} and {\sc Peres} \cite{Per3}). We shall 
discuss this problem from our perspective in a later part of this 
sequence of papers.

\section{Dynamics} \label{dynamics}

\hfill\parbox[t]{6.5cm}{\footnotesize

{\em The lot is cast into the lap; 
but its every decision is from the {\sc LORD}.}

King Solomon, ca. 1000 B.C. \cite{Sol} 

\bigskip
{\em God does not play dice with the universe.}

Albert Einstein, 1927 A.D. \cite{Ein}
}\nopagebreak

\bigskip
In this section we discuss the most elementary aspects of the dynamics 
of (closed and isolated) physical systems. 
We shall have much more to say about dynamics in 
later parts of this series of papers, where so-called Poisson algebras
will be used to make the formal dynamical parallels between classical 
mechanics and quantum mechanics understandable as two special cases of 
a single theory.

The observations about a physical system change with time. The dynamics 
of a closed and isolated physical system is conservative, and may be
described by a fixed (but system-dependent) 
one-parameter family $S_t$ ($t\in\Rz$) of {\bf automorphisms} of the 
*-algebra $\Ez$, i.e., mappings $S_t:\Ez\to\Ez$ satisfying 
(for $f,g \in \Ez$, $\alpha\in\Cz$, $s,t\in\Rz$)

(A1) 
~{$S_t(\alpha)=\alpha, ~~~ S_t(f^*)=S_t(f)^*$,}

(A2)
~{$S_t(f+g)=S_t(f)+S_t(g), ~~~ S_t(fg)=S_t(f)S_t(g)$,}

(A3)
~{$S_0(f)=f, ~~~ S_{s+t}(f)=S_s(S_t(f))$.}

In the {\bf Heisenberg picture} of the dynamics, where states are
fixed and quantities change with time, $f(t):=S_t(f)$ denotes the 
time-dependent {\bf Hei\-senberg quantity} associated with 
$f$ at time $t$. Note that $f(t)$ is uniquely determined by $f(0)=f$. 
Thus the dynamics is deterministic, {\em independent of whether we 
are in a classical or in a quantum setting}.

(In contrast, nonisolated closed systems are dissipative and 
intrinsically sto\-chastic; see, e.g., 
{\sc Giulini} et al. \cite{GiuJK}.
We shall discuss this in a later part of this series.)

\begin{expls} \label{ex.classquant.dyn}
In nonrelativistic mechanics, conservative systems are described by a 
Hermitian quantity $H$, called the {\bf Hamiltonian}.

(i) In {\bf classical mechanics} 
-- cf. Example \ref{ex.classquant}(i) --,
a Poisson bracket $\{\cdot,\cdot\}$ together with $H$ defines the 
Liouville superoperator $Lf=\{f,H\}$, and the dynamics is given by 
the one-parameter group defined by
\[
S_t(f)=e^{tL}(f),
\]
corresponding to the differential equation
\lbeq{e.heisc}
\frac{df(t)}{dt}=\{f(t),H\}.
\eeq

(ii) In {\bf nonrelativistic quantum mechanics} 
-- cf. Example \ref{ex.classquant}(ii) --, 
the dynamics is given by the one-parameter group defined by
\[
S_t(f)=e^{-tH/i\hbar}fe^{tH/i\hbar},
\] 
corresponding to the {\bf Heisenberg equation}
\lbeq{e.heisq}
i\hbar\frac{df(t)}{dt}=e^{-tH/i\hbar}[f,H]e^{tH/i\hbar}=[f(t),H].
\eeq

(iii) {\bf Relativistic quantum mechanics} is currently 
(for interacting systems)
developed only for scattering events in which the dynamics is 
restricted to transforming quantities of a system at $t=-\infty$ to 
those at $t=+\infty$ by means of a single automorphism $S$ given by
\[
S(f)=sfs^*,
\]
where $s$ is a unitary quantity (i.e., $ss^*=s^*s=1$), the so-called 
{\bf scattering matrix}, for which an asymptotic series in powers of
$\hbar$ is computable from quantum field theory.

\end{expls}

The realization of the axioms is different in the classical and 
in the quantum case, but the interpretation is identical. 

The common form and deterministic nature of the dynamics, independent 
of any assumption of whether the system is classical or quantum, 
implies that there is no difference in the causality of classical 
mechanics and that of quantum mechanics. Therefore, 
{\em the differences between classical mechanics and quantum mechanics 
cannot lie in an assumed intrinsic indeterminacy of quantum mechanics 
contrasted to deterministic classical mechanics}. 
The only difference between classical mechanics and quantum mechanics 
in the latter's lack of commutativity.

\bigskip
Of course, reference values of quantities at different times 
will generally be different. To see what happens, suppose that,
in state $v$, a quantity $f$ has reference value $v(f)$ at time 
$t=0$. At time $t$, the quantity $f$ developed into $f(t)$, with 
reference value 
\lbeq{e.SH}
v(f(t))=v(S_t(f))=v_t(f),
\eeq
where the time-dependent {\bf Schr\"odinger state}
\lbeq{e.schro}
v_t=v\circ S_t
\eeq
is the composition of the two mappings $v$ and $S_t$.
It is easy to see that $v_t$ is again a state, and that all properties 
discussed in the previous section that $v$ may possess are inherited 
by $v_t$.

Thus we may recast the dynamics in the {\bf Schr\"odinger picture}, 
where quantities are fixed and states change with time. The dynamics
of the time-dependent states $v_t$ is then given by \gzit{e.schro}. 
Of course, in this picture, the dynamics is deterministic, too.

\begin{expls}~

(i) In {\bf classical mechanics}, \gzit{e.heisc} implies
for an ensemble state of the form 
\[
v_t(f)=\int_{\Omega_{cl}} \rho(\omega,t)f(\omega)d\omega
\]
the {\bf Liouville equation}
\[
i\hbar\frac{d\rho(t)}{dt}=\{H,\rho(t)\}.
\]

(ii) In {\bf nonrelativistic quantum mechanics}, \gzit{e.heisq} implies
for an ensemble state of the form 
\[
v_t(f)=\tr \rho(t)f 
\]
the {\bf von Neumann equation}
\[
i\hbar\frac{d\rho(t)}{dt}=[H,\rho(t)].
\]

(iii) {\bf Bohmian mechanics} has no natural Heisenberg picture,
cf. {sc Holland} \cite[footnote p.519]{Hol}. The reason is that
noncommuting position operators at different times are assumed 
to have sharp values. Thus the results of this section do not apply 
to it. 

\end{expls}

\bigskip
In a famous paper, {\sc Einstein, Podolsky \& Rosen} \cite{EinPR} 
introduced the following criterion for elements of physical reality:

{\em 
If, without in any way disturbing a system, we can predict with 
certainty (i.e., with probability equal to unity) the value of a 
physical quantity, then there exists an element of physical reality 
corresponding to this physical quantity
}

and postulated that

{\em 
the following requirement for a complete theory seems to be a 
necessary one: every element of the physical reality must have a 
counterpart in the physical theory.
}

Traditionally, elements of physical reality were thought to have to
emerge in a classical framework with hidden variables.
However, to embed quantum mechanics in such a framework is impossible 
under natural hypotheses ({\sc Kochen \& Specker} \cite{KocS});
indeed, it amounts to having states in which all Hermitian quantities
are sharp, and we have seen that this is impossible for quantum
systems involving a Hilbert space of dimension $4$ or more.
 
However, the reference values of states with numerical reference 
values for {\em all} Hermitian quantities, and in particular the
reference values of ensemble states,
are such elements of physical reality: 
If one knows in a state $v=v_0$ all reference values with 
certainty at time $t=0$ then, since the dynamics is deterministic, one 
knows with certainty the reference values \gzit{e.SH} at any time.
In this sense, ensemble states provide a realistic interpretation of 
quantum mechanics.

\bigskip
Taking another look at the form of the Schr\"odinger dynamics
\gzit{e.SH}, we see 
that the reference values behave 
just like the particles in an ideal fluid, propagating independently 
of each other. We may therefore say that 
the Schr\"odinger dynamics describes the {\bf flow of truth}
in an objective, deterministic manner. On the other hand,
the Schr\"odinger dynamics is completely silent 
about {\em what} is true. Thus, as in mathematics, where all truth is 
relative to the logical assumptions made (what is considered true at 
the beginning of an argument), in theoretical physics truth is relative 
to the initial values assumed (what is considered true at time $t=0$).

In both cases, theory is about what is consistent, and not about what 
is real or true. The formalism enables us only to deduce truth from 
other assumed truths. But what is regarded as true is outside the 
formalism, may be quite subjective and may even turn out to be 
contradictory, depending on the acquired personal habits of 
self-critical judgment. 

What we can possibly know as true are the {\em laws} of physics, 
general relationships that appear often enough to see the underlying 
principle. But concerning {\em states} (i.e., in practice, boundary 
conditions) we are doomed to idealized, more or less inaccurate 
approximations of reality. {\sc Wigner} \cite[p.5]{Wig} expressed
this by saying,
{\em the laws of nature are all conditional statements and they relate 
only to a very small part of our knowledge of the world.}

\bigskip
\section{Epilogue} 

The axiomatic foundation given here of the basic principles underlying 
theoretical physics suggests that, from a formal point of view, the 
differences between classical physics and quantum physics are only 
marginal (though in the quantum case, the lack of commutativity 
requires some care and causes deviations from classical behavior). 
In both cases, everything flows from the same assumptions simply by 
changing the realization of the axioms.

It is remarkable that, in the setting of Poisson algebras described 
and explored in later parts of this series of papers, this remains so 
even as we go deeper into the details of dynamics and thermodynamics.

\bigskip


\begin{thebibliography}{99}

\bibitem{Asp} A. Aspect,
Proposed experiment to test the nonseparability of quantum mechanics,
Phys. Rev. D 14 (1976), 1944-1951.
(Reprinted in \cite{WheZ}.)

\bibitem{Bel} J.S. Bell,
On the Einstein Podolsky Rosen paradox,
Physics 1 (1964), 195-200.
(Reprinted in \cite{WheZ}.)

\bibitem{Bel2} J.S. Bell,
Speakable and unspeakable in quantum mechanics,
Cambridge Univ. Press, Cambridge 1987.

\bibitem{Ber} H.J. Bernstein,
Simple version of the Greenberger-Horne-Zeilinger (GHZ) argument 
against local realism,
Found. Phys. 29 (1999), 521-525.

\bibitem{BirN} G. Birkhoff and J. von Neumann,
The logics of quantum mechanics,
Ann. Math. 37 (1936), 823-843.

\bibitem{Boh} D. Bohm,
A suggested interpretation of the quantum theory in terms of `hidden'
variables, I and II,
Phys. Rev. 85 (1952), 166-179.
(Reprinted in \cite{WheZ}.)

\bibitem{Bohr} N. Bohr,
Discussion with Einstein on epistemological problems in atomic physics,
pp. 200-241 in: 
P.A. Schilpp (ed.), 
Albert Einstein: Philosopher-Scientist,
The Library of Living Philosophers, Evanston 1949.
(Reprinted in \cite{WheZ}.)

\bibitem{BusGL}
P. Busch, M. Grabowski and P.J. Lahti,
Operational quantum physics,
Springer, Berlin 1995.

\bibitem{BusLM}
P. Busch, P.J. Lahti and P. Mittelstaedt,
The quantum theory of measurement, 2nd. ed.,
Springer, Berlin 1996.

\bibitem{Cir} B.S. Cirel'son, 
Quantum generalizations of Bell's inequality, 
Lett. Math. Phys. 4 (1980), 93-100.

\bibitem{ClaHS} J. F. Clauser, M.A. Horne, A. Shimony and R.A. Holt,
Proposed experiment to test local hidden-variable theories,
Phys. Rev. Lett. 23 (1969), 880-884.
(Reprinted in \cite{WheZ}.)

\bibitem{ClaS} J. F. Clauser and A. Shimony,
Bell's theorem: experimental tests and implications,
Rep. Prog. Phys. 41, 1881-1926 (1978).

\bibitem{CliK} R. Clifton and A. Kent,
Simulating quantum mechanics by non-contextual hidden variables,
Manuscript (1999). quant-ph/9908031.

\bibitem{Dav} E.B. Davies,
Quantum theory of open systems,
Academic Press, London 1976.

\bibitem{Dir} P.A.M. Dirac, 
Lectures on quantum field theory. 
Belfer Grad. School of Sci., New York 1966.

\bibitem{Dri} M. Drieschner, 
Voraussage -- Wahrscheinlichkeit -- Objekt. 
\"Uber die begrifflichen Grundlagen der Quantenmechanik.
Lecture Notes in Physics, Springer, Berlin, 1979.

\bibitem{Ebe} P.H. Eberhard,
Bell's theorem without hidden variables,
Il Nuovo Cimento 38 B (1977), 75-80.

\bibitem{Ein} A. Einstein, 
Conversation with Bohr and Ehrenfest at the Fifth Solvay conference in 
October, 1927; cf. \newl
{\small\tt http://solon.cma.univie.ac.at/\wave neum/contrib/dice.txt}
\newl
The formulation used is from a letter of September 7, 1944, 
reprinted pp. 275-276 in: 
A.P. French (ed.), 
Einstein, a centenary volume,
Harvard Univ. Press, Cambridge, Mass. 1979.

\bibitem{Ein2} A. Einstein, 
Einleitende Bemerkungen \"uber Grundbegriffe, in:
Louis de Broglie, physicien et penseur (A. George, ed.),
Albin Michel, Paris 1953.
quoted after \cite{dEs}, p. 407.

\bibitem{EinPR} A. Einstein, B. Podolsky and N. Rosen,
Can the quantum-mechanical description of physical reality be
considered complete?
Phys. Rev. 47 (1935), 777-780.
(Reprinted in \cite{WheZ}.)

\bibitem{dEs} B. d'Espagnat,
Veiled reality. An analysis of present-day quantum mechanical concepts,
Addison-Wesley, Reading, Mass., 1995.

\bibitem{Eve} H. Everett III, 
Relative state formulation of quantum mechanics, 
Rev. Mod. Phys. 29 (1957) 454-462.
(Reprinted in \cite{WheZ}.)

\bibitem{Fin} T.L. Fine, 
Theory of probability; an examination of foundations. 
Acad. Press, New York 1973.

\bibitem{GiuJK} D. Giulini, E. Joos, C. Kiefer, J. Kupsch, 
I.-O. Stamatescu and H.D. Zeh,
Decoherence and the appearance of a classical world in quantum theory, 
Springer, Berlin 1996.

\bibitem{GreHS} D.M. Greenberger, M.A. Horne, A. Shimony and 
A. Zeilinger,
Bell's theorem without inequalities,
Amer. J. Phys. 58 (1990), 1131-1143.

\bibitem{GreHZ} D.M. Greenberger, M.A. Horne and A. Zeilinger,
Going beyond Bell's theorem,
pp. 73-76 in:
M. Kafatos (ed.), 
Bell's theorem, quantum theory, and conceptions of the universe,
Kluwer, Dordrecht 1989.

\bibitem{Hac} I. Hacking,
The emergence of probability,
Cambridge Univ. Press, Cambridge 1975.

\bibitem{Har} L. Hardy,
Quantum mechanics, local realistic theories, and Lorentz-invariant 
realistic theories,
Phys. Rev. Lett. 68 (1992), 2981-2984.

\bibitem{Har2} L. Hardy,
Nonlocality for two particles without inequalities for almost all 
entangled states,
Phys. Rev. Lett. 71 (1993), 1665-1668.

\bibitem{Haw} M. Hawton, 
Photon position operator with commuting components, 
Phys. Rev. A 59 (1999), 954-959. 

\bibitem{Hei} W. Heisenberg,
{\"U}ber den anschaulichen Inhalt der quantentheoretischen Kinematik 
und Mechanik,
Zeitschrift f. Physik 43 (1927), 172-198.
(Engl. translation: Section I.3 in \cite{WheZ}.)

\bibitem{Hig} N.J. Higham,
Accuracy and stability of numerical algorithms,
SIAM, Philadelphia 1996.

\bibitem{Hil} D. Hilbert, 
Mathematische Probleme, 
Bull. Amer. Math. Soc. 8 (1902), 437-479.

\bibitem{HofMC} R. Hoffmann, V.I. Minkin and B.K. Carpenter,
Ockham's Razor and Chemistry,
HYLE Int. J. Phil. Chem 3 (1997), 3-28.\newl
{\small \tt http://rz70.rz.uni-karlsruhe.de/~ed01/Hyle/Hyle3/hoffman.htm}
\bibitem{Hol} P.R. Holland, 
The quantum theory of motion,
Cambridge Univ. Press, Cambridge 1993.

\bibitem{HomW} D. Home and M.A.B. Whitaker,
Ensemble interpretations of quantum mechanics. A modern perspective,
Phys. Rep. 210 (1992), 223-317.

\bibitem{Isa} Isaiah 55:9,
Holy Bible, New International Version, 1984.

\bibitem{Isa2} Isaiah 65:24,
Holy Bible, New International Version, 1984.

\bibitem{Jam1} M. Jammer,
The conceptual development of quantum mechanics,
McGraw-Hill, New York 1966.

\bibitem{Jam2} M. Jammer,
The philosophy of quantum mechanics: 
the interpretations of quantum mechanics in historical perspective,
Wiley, New York 1974.

\bibitem{Jau} J.M. Jauch, 
Foundations of quantum mechanics,
Addison-Wesley, Reading, MA 1968.

\bibitem{KocS} S. Kochen and E.P. Specker,
The problem of hidden variables in quantum mechanics,
J. Math. Mech. 17 (1967), 59-67.
(Reprinted in C.A. Hooker, ed.,
The logico-algebraic approach to quantum mechanics,
Vol. I: Historical evolution,
Reidel, Dordrecht 1975.)

\bibitem{Koh2} Kohelet, Ecclesiastes 1:10,
in: Holy Bible, Good News Edition, 1984.

\bibitem{Koh} Kohelet, Ecclesiastes 11:6,
in: Holy Bible, New International Version, 1984.

\bibitem{Kol} A.N. Kolmogorov,
Foundations of the theory of probability,
Chelsea, New York 1950.
(German original: 
Grundbegriffe der Wahr\-schein\-lich\-keits\-rech\-nung,
Sprin\-ger, Berlin 1933.)

\bibitem{ManW} L. Mandel and E. Wolf,
Optical coherence and quantum optics,
Cambridge Univ. Press 1995.

\bibitem{Mer} N.D. Mermin,
Simple unified form for the major no-hidden-variables theorems,
Phys. Rev. Lett. 65 (1990), 3373-3376.

\bibitem{Mer2} N.D. Mermin,
What's wrong with these elements of reality?
Physics Today (June 1990), 9-10.

\bibitem{Mes} A. Messiah,
Quantum mechanics, Vol. 1,
North-Holland, Amsterdam 1991;
Vol. 2,
North-Holland, Amsterdam 1976.

\bibitem{Mit} P. Mittelstaedt,
The interpretation of quantum mechanics and the measurement process,
Cambridge Univ. Press, Cambridge 1998.

\bibitem{Mey} P.-A. Meyer,
Quantum probability for probabilists, 2nd. ed., 
Springer, Berlin 1995.

\bibitem{Neu.bohm} A. Neumaier,
Bohmian mechanics contradicts quantum mechanics,
Manuscript (2000).
quant-ph/0001011

\bibitem{Neu.prot} A. Neumaier,
Molecular modeling of proteins and mathematical prediction of 
protein structure,
SIAM Rev. 39 (1997), 407-460.

\bibitem{vNeu} J. von Neumann,
Mathematische Grundlagen der Quantenmechanik.
Springer, Berlin 1932.

\bibitem{NewW}
T.D. Newton and E.P. Wigner,
Localized states for elementary systems,
Rev. Mod. Phys 21 (1949), 400-406.

\bibitem{Ock} W. of Ockham,
Philosophical Writings, 
(ed. by P. Boehner)
Nelson, Edinburgh 1957. 

\bibitem{Par} K.R. Parthasarathy,
An introduction to quantum stochastic calculus,
Birkh\"auser, Basel 1992.

\bibitem{Pau} St. Paul, 1 Corinthian 13:8-10,
in: The New Testament. This is my paraphrase 
of a famous quote by Paul; for other renderings, see, e.g.,\newl
{\small\tt http://solon.cma.univie.ac.at/\wave neum/christ/contrib/1cor13.html}

\bibitem{Pau3} St. Paul, 1 Timothy 6:8,
in: Holy Bible, New International Version, 1984.

\bibitem{Per} A. Peres,
In compatible results of quantum measurements,
Physics Lett. A 151 (1990), 107-108.

\bibitem{Per2} A. Peres,
Two simple proofs of the Kochen-Specker theorem,
J. Phys. A: Math. Gen. 24 (1991), L175-L178.

\bibitem{Per3} A. Peres, 
Quantum theory: Concepts and methods,
Kluwer, Dordrecht 1993.

\bibitem{Pit} I. Pitowsky,
Quantum probability -- quantum logic,
Lecture Notes in Physics 321,
Springer, Berlin 1989.

\bibitem{Pla} Plato, 
Timaeus,
Hackett Publishing, Indianapolis 1999. The quotes (Tim. 28-29) are from the 
Project Gutenberg Etext at
\newl
{\footnotesize\tt ftp://metalab.unc.edu/pub/docs/books/gutenberg/etext98/tmeus11.txt}\newl
(the first half of the document is a commentary, then follows the 
original in English translation)

\bibitem{Pry}
M.H.L. Pryce, 
The mass-centre in the restricted theory of relativity and its 
connexion with the quantum theory of elementary particles,
Proc. Roy. Soc. London A 195 (1949), 62-81.

\bibitem{Ric} C.E. Rickart,
General theory of Banach algebras.
Van Nostrand, Princeton 1960.

\bibitem{Rob} H.P. Robertson,
The uncertainty principle,
Phys. Rev. 34 (1929), 163-164.
(Reprinted in \cite{WheZ}.)

\bibitem{Rou} T. Roubicek,
Relaxation in Optimization Theory and Variational Calculus, 
Walter de Gruyter, Berlin 1997.

\bibitem{Sch} E. Schr{\"o}dinger,
Die gegenw{\"a}rtige Situation in der Quantenmechanik, 
Naturwissenschaften 23 (1935), 807-812; 823-828; 844-849.
(Engl. translation: Section I.11 in \cite{WheZ}.)

\bibitem{Sol} King Solomon, Proverbs 16:3,
in: Holy Bible, New International Version, 1984.

\bibitem{Sto} M.H. Stone,
The theory of representations for Boolean algebras, 
Trans. Amer. Math. Soc. 40 (1936), 37-111.

\bibitem{Str} J. Strnad, 
Photons in introductory quantum physics,
Amer. J. Phys. 54 (1986), 650-652.

\bibitem{Svo} K. Svozil,
Quantum logic. 
Springer, Berlin 1998.

\bibitem{TitBG} W. Tittel, J. Brendel, B. Gisin, T. Herzog, H. Zbinden 
and N. Gisin,
Experimental demonstration of quantum-correlations over more than 
10 kilometers,
Phys. Rev. A 57, 3229-3232 (1998). 

\bibitem{Vai} L. Vaidman,
Variations on the theme of the Greenberger-Horne-Zeilinger proof,
Found. Phys. 29 (1999), 615-630.

\bibitem{WheZ} J.A. Wheeler and W. H. Zurek,
Quantum theory and measurement.
Princeton Univ. Press, Princeton 1983.

\bibitem{Whi} P. Whittle,
Probability via expectation, 3rd ed.,
Springer, New York 1992.
(1st ed.: Probability, Harmondsworth 1970.)

\bibitem{Wig} E.P. Wigner,
The unreasonable effectiveness of mathematics in the natural sciences,
Comm. Pure Appl. Math. 13 (1960), 1-14.

\bibitem{WooZ} W.K. Wootters and W.H. Zurek,
Complementarity in the double-slit experiment: Quantum nonseparability 
and a quantitative statement of Bohr's principle,
Phys. Rev. D 19 (1979), 473-484.
(Reprinted in \cite{WheZ}.)

\bibitem{Zim} H.-J. Zimmermann, 
Fuzzy set theory -- and its applications, 3rd ed.,
Kluwer, Dordrecht 1996. 

\end{thebibliography}
\end{document}